\documentclass{ws-ijmpe}

\usepackage{dcolumn}
\usepackage{bm}
\usepackage{amsmath}
\usepackage{amssymb}

\newcommand{\nn}{\nonumber}

\newcommand{\beq}{\begin{equation}}
\newcommand{\eeq}{\end{equation}}
\newcommand{\beqa}{\begin{eqnarray}}
\newcommand{\eeqa}{\end{eqnarray}}

\newcommand{\fracpd}[2]{\frac{\partial #1}{\partial #2}} 
\newcommand{\g}{g_{\mu\nu}}
\newcommand{\dd}{\mathrm{d}}

\begin{document}
\def\ii{\'\i}

\markboth{G. Caspar, T. Sch\"onenbach, P. O. Hess, M. Sch\"afer, W. Greiner}{Pseudo-complex General Relativity: Schwarzschild, Reissner-Nordstr\"om
and Kerr solutions}

\catchline{}{}{}{}{}

\title{PSEUDO-COMPLEX GENERAL RELATIVITY:\\
SCHWARZSCHILD, REISSNER-NORDSTR\"OM AND KERR SOLUTIONS
}

\author{\footnotesize GUNTHER CASPAR}

\address{Frankfurt Institute for Advanced Studies, Johann Wolfgang Goethe University, Ruth-Moufang-Str. 1  \\
Frankfurt am Main, 60438,
Germany
\\
caspar@th.physik.uni-frankfurt.de}

\author{THOMAS SCH\"ONENBACH}

\address{Frankfurt Institute for Advanced Studies, Johann Wolfgang Goethe University, Ruth-Moufang-Str. 1  \\
Frankfurt am Main, 60438,
Germany
\\
schoenenbach@fias.uni-frankfurt.de}

\author{PETER OTTO HESS}

\address{Instituto de Ciencias Nucleares, UNAM, Circuito Exterior, C.U.  \\
A.P. 70-543, 04510 M\'exico D.F. , Mexico
\\
hess@nucleares.unam.mx}

\author{MIRKO SCH\"AFER}

\address{Frankfurt Institute for Advanced Studies, Johann Wolfgang Goethe University, Ruth-Moufang-Str. 1  \\
Frankfurt am Main, 60438,
Germany
\\
schaefer@fias.uni-frankfurt.de}

\author{WALTER GREINER}

\address{Frankfurt Institute for Advanced Studies, Johann Wolfgang Goethe University, Ruth-Moufang-Str. 1  \\
Frankfurt am Main, 60438,
Germany
\\
greiner@fias.uni-frankfurt.de}

\maketitle

\begin{history}
\received{(received date)}
\revised{(revised date)}
\end{history}

\begin{abstract}
The pseudo-complex General Relativity (pc-GR) is further considered. A new
projection method is proposed.
It is shown, that the pc-GR introduces automatically terms into the system
which can be interpreted as dark energy. The modified pseudo-complex
Schwarzschild solution is investigated. The dark energy part is treated as a
liquid and possible solutions are discussed. As a consequence, the collapse
of a large stellar mass into a singularity at $r=0$ is avoided and no
event-horizon is formed. Thus, black holes don't exist. The resulting object 
can be viewed as a gray star. It contains no singularity which emphasizes, again, 
that it is not a black hole. The corrections implied
by a charged large mass object (Reissner- Nordstr\"om) and a rotating
gray star (Kerr) are presented.
For the latter, a special solution is presented.
Finally, we will consider the orbital speed of a mass in a circular orbit and suggest
 a possible experimental verification.
\end{abstract}
\vskip 0.5cm
\noindent
PACS: 02.40.ky, 98.80.-k

\vskip 1cm

\section{Introduction}

General Relativity (GR) is an extremely successful theory, which up to now
is confirmed by a great number of experiments. Nevertheless, it contains a
singularity for highly dense small objects (Schwarzschild singularity) 
and the appearance of singularities in any field theory shows, that its 
correct form has not yet been found.
Einstein's GR predicts the existence of {\it black holes},
which imply singularities at the center of great masses and the appearance of
an event horizon at the Schwarzschild radius, $r_S$. Due to this reason (and some
others which we will mention later on) several groups tried to extend
GR such that the new theory does not contain the mentioned
undesired property.

In an attempt to extend GR, for example, Einstein in \cite{einstein1,einstein2} substituted the real
space-time variables by complex ones, an idea which was continued in
\cite{mantz,lovelook} and called it {\it complex GR}. Others took the idea
of M. Born \cite{born1,born2}, which treats the space-time coordinates and 
its conjugate momenta  on equal footing. Indeed in Quantum
Mechanics coordinates and momenta can be interchanged through a canonical transformation.
Thus, there might be hope to connect GR to Quantum Mechanics.
Our proposal consists in the redefinition of the length square element,
which approaches the one of standard GR for large masses and distances, while
for small distances the coordinates and momenta appear
symmetrically in the length square element.
This extension
introduces a maximal acceleration, as was shown in \cite{caneloni}. In
\cite{brandt1,brandt2,brandt3,beil1,beil2,beil3,beil4,moffat1,moffat2,kunstatter}
the consequences of this approach were investigated. In another
attempt to extend GR, hyberbolic (another name for pseudo-complex) coordinates were introduced
\cite{crumeyrolle1,crumeyrolle2,clerc1,clerc2}.
In \cite{hess1} we extended GR to pseudo-complex coordinates. The new pseudo-complex General Relativity (pc-GR) contains 
the former mentioned theories. The pseudo-imaginary component is associated with
the components of the tangent vector (four momentum) at a given space-time point.
Due to dimensional reasons, a minimal
length scale ($l$) is introduced into the theory. It corresponds to a
maximal acceleration. The pc-Schwarzschild solution was obtained, 
neglecting all corrections due to the minimal length
scale $l$.
Due to the pc-description an additional contribution appears in the energy-momentum
tensor. One can say, that the modified field equations can be interpreted
(within the old Einsteinian GR) as dark energy: The additional terms play the role of a repulsive dark energy.
The energy accumulates around a central mass and halts the collapse of a
big star. No singularity and no event horizon forms. Thus,
no black hole exists in this theory. This is an important achievement!
The new highly dense objects may be called
{\it gray stars}.
In \cite{hess2} the pc-Robertson-Walker model of the universe
was proposed. Also there, terms, which can be interpreted as ``dark energy'' appear automatically, 
and new solutions are obtained, like
a universe which at large time tends to a constant acceleration or
diminishes its acceleration, approaching zero.

In this paper we refer for an introduction to pseudo-complex
variables to \cite{hess1,hess2} and \cite{hess3}.
An extensive mathematical treatment of pseudo-complex coordinates can be found in \cite{anton,kantor}.

We revisit the initially proposed pseudo-complex
extension of GR in \cite{hess1,hess2}, where the projection method suffered
from conceptual difficulties: for example the projected metric
did not represent a tensor. In section 2 we will investigate an alternative
projection method which will result in a real metric with tensorial
properties.
In section 3 we will apply it to the pc-Schwarzschild solution.
As a new ingredient, the dark energy is treated as a fluid and 
the differential equations are derived in sub-section 3.3. Both the pressure and
density have to obey these equations.
Some approximate solutions are discussed. In section 4
a charged gray star, i.e. the pc-Reissner-Nordstr\"om solution,
is investigated.
Section 5 deals with the pc-Kerr problem.
The equations of motion will be formulated.
A specific solution will also be discussed, which shows similarities to the
standard solution of Einsteins GR, with the essential difference that, again, black holes
do not exists.
The results can
be used for further studies to find possible analytic solutions for
a rotating gray star.

We also discuss in section 6 possible experimental verifications for further testing
our theory. In particular the orbital speed of a mass in a circular orbit around a gray star is
considered.

Section 7 contains the conclusions.

\section{Reformulation of the pseudo-complex General Relativity}

Let us first briefly remind on some key properties of pc-variables:
A pc coordinate $X$ is given by $X=X_R+I X_I$, where $X_R$
is the pseudo-real and $X_I$ its pseudo-imaginary component. Contrary to
complex variables, the $I^2=1$. This is, at first, astonishing. But remember that e.g. 
for the Pauli-matrices $\hat{\sigma}_i$ one has $\hat{\sigma}_i^2 = 1$. Thus one can
think of pseudo-complex coordinates as matrix-coordinates. Have in mind
a similar situation, when Dirac introduced instead of 
$E=\pm\sqrt{\left ( pc\right )^2 + \left ( m_0c^2\right )^2}$ the matrix
$E= \sum_i \gamma^i \hat{p}_i$. This lead from the Klein-Gordon equation
to the Dirac equation, which contained spin and was the basis for a model
of the vacuum and for creation of particle-antiparticle pairs.  

A pseudo-complex conjugate variable is
defined as $X^*=X_R - I X_I$. Instead of this representation one can express
the variables as a linear combination of $\sigma_\pm = \frac{1}{2}
\left( 1 \pm I\right)$, i.e., $X=X_+\sigma_+ + X_-\sigma_-$, with
$X_{R/I} = \frac{1}{2}\left( X_+ \pm X_-\right)$ and $\sigma_\pm^2 = \sigma_\pm$,
$\sigma_+ \sigma_- = 0$. The last property implies that the pseudo-complex
variables
contain a zero divisor \cite{anton,kantor}, which produces very important changes in
the structure of GR. Variables, which are within the zero divisor,
satisfy the property $|X|^2=X^* X=0$. These variables can be expressed either as
$X=X_+ \sigma_+$ or $X=X_- \sigma_-$ and they constitute the
{\it zero divisor}
denoted by ${\cal P}^0$. One important property of the use of pseudo-complex
variables is that {\it any mathematical application can be done independently
within the $\sigma_+$ and the $\sigma_-$ component}. For example, the
multiplication of two functions $F(X)=F_+(X_+)\sigma_+ + F_-(X_-)\sigma_-$
and $G(X)=G_+(X_+)\sigma_+ + G_-(X_-)\sigma_-$ is just
$F(X)G(X)=F_+(X_+)G_+(X_+)\sigma_+ + F_-(X_-)G_-(X_-)\sigma_-$. This
property will enable us later to define two different metric tensors, one
in the $\sigma_+$ and the other one in the $\sigma_-$ sector and that in each
sector we can define a standard GR.
The $X_\pm$ will be denoted as the {\it zero-divisor components} of $X$.

In a pc-theory, the variational principle has to be modified, such that

\beqa
\delta S & \epsilon & {\cal P}^0
~~~,
\eeqa
where ${\cal P}^0$ represents the zero divisor.
One can show \cite{hess1,hess3} that this change in the variational
principle is necessary because otherwise one would generate
two independent theories, one in the $\sigma_+$ and the other one in the
$\sigma_-$ sector of the action. In a pc-theory one has then
to worry on how to connect both components, when real results
are projected out.

For convenience we choose the variation of the action
proportional to $\sigma_-$ and note that using $\sigma_+$ results in the
same description: only the $\sigma_+$ with the $\sigma_-$
component of the theory are interchanged.

Extending a theory to a pseudo-complex version will, in general, introduce
important simplifications.
For example, as shown in
\cite{hess3,schuller0,schuller1} the pseudo-complex extension of
a {\it linear} Field Theory 
leads to a regularized theory \'a la Pauli-Villars, {\it maintaining
the linear description} of the field theory, while not applying the pc-extension
the Lagrange density is highly non-linear.

In a pc-field theory, the projection to the real part
is obtained by noting that the pseudo-complex
extension of the Lorentz group (similar for the Poincar\'e group) is
given by \cite{field1,field2}

\beqa
SO_+(3,1) \otimes SO_-(3,1) \supset SO(3,1)
~~~,
\eeqa
where finite transformations in the $SO_+(3,1) \otimes SO_-(3,1)$ groups
and the projection to $SO(3,1)$ are given by

\beqa
e^{\omega^{\mu\nu} L_{\mu\nu}} \rightarrow e^{\omega^{\mu\nu}_R L^R_{\mu\nu}}
~~~.
\label{projection}
\eeqa
Here, $\omega^{\mu\nu}=\omega^{\mu\nu}_R + I \omega^{\mu\nu}_I$
are the pseudo-complex transformation parameters and
$L_{\mu\nu} = X_\mu P_\nu - X_\nu P_\mu$ are the pseudo-complex
generators.
In the projection of $L_{\mu\nu}$ to $L_{\mu\nu}^R$,
the coordinates $X_\mu$ are substituted by $x_\mu$ and
the momenta $P_\nu$ by $p_\nu$.
The finite transformation is a function
$f(\omega_{\mu\nu} ,X_\mu , P_\mu)$ in the pseudo-complex coordinates
($X_\mu$), momenta ($P_\mu$) and transformation parameters
($\omega_{\mu\nu}$). As can be seen, the function $f$ is not mapped to
$\frac{1}{2}\left( f + f^* \right)$ =
$\frac{1}{2}\left( f_+ + f_- \right)$, which is also real but not necessary
the same function anymore, but rather
the {\it arguments} of the function are mapped to their real values,
yielding {\it the same function} but with real arguments.
To illustrate that, we consider the example of a function $f= \frac{1}{R + A}$,
where $A$ has no $\sigma_+$-part, e.g. $A = A_- \sigma_-$. $\frac{1}{2}\left( f_+ + f_- \right)$
is then given by
\begin{equation}
 \frac{1}{2}\left( f_+ + f_- \right) =  \frac{1}{2}\left( \frac{1}{R_+ } + \frac{1}{R_- + A_-} \right) = \frac{1}{2}\left( \frac{R_- + A_- +R_+}{R_+(R_- +A_-)} \right) = \frac{r + A_\text{Re}}{r^2 + 2  r A_\text{Re}} \quad .
\end{equation}
On the other hand, if we map the arguments of $f$ we recieve
\begin{equation}
 f= \frac{1}{R + A} \rightarrow \frac{1}{r + A_\text{Re}}
\end{equation}

This example has to be kept in mind, when we
define the projection to the real space in the pc-extension of GR, i.e.,
the proposed projection will substitute in any function of the
pseudo-complex coordinates, momenta and parameter values to
{\it the same function} but now in the real coordinates, momenta and
parameter values.

There is another simplification provided by the pc-description:
As shown in \cite{hess3}, the field equation for a 
scalar boson field is obtained from the Lagrangian density
$\frac{1}{2}\left( D_\mu \Phi D^\mu \Phi - M^2 \Phi^2 \right)$, where
$\Phi$ is the pc-boson field, $M=M_+\sigma_+ + M_-\sigma_-$
is a pc-mass and $D_\mu$ a pc-derivative.
The propagators of this theory are the ones of Pauli-Villars,
which already are regularized. One obtains the same propagator in the standard
theory, with non-pc scalar field, using the Lagrange density
$-\frac{1}{\left(M_+^2 - M_-^2\right)} \phi
\left( \partial_\mu\partial^\mu + M_+^2 \right)$ 
$\left( \partial_\mu\partial^\mu + M_-^2 \right)\phi$, where
$\phi$ is now a real valued function,
$M_+$ is identified with the physical mass $m$ and $M_->>M_+$ with the
regularizing mass. Note, that this theory is highly non-linear while
the pc-description is linear. This indicates that a pc-description can
substantially simplify the structure of the theory.
%

In \cite{moffat1,moffat2} Moffat el al. proposed a possible extension 
of GR by introducing a non-symmetric metric. 
In \cite{ghosts} the symmetric part of the metric
was associated to the pseudo-real component of a pseudo-complex metric
(called in \cite{ghosts} hyper-complex) and the antisymmetric metric was
associated to the pseudo-imaginary component. Investigating the linearized
limit for small fields, it was shown that the Lagrangian separates
into two parts, one depending only on the symmetric and the other one on the
anti-symmetric components. It was proven that
due to $I^2=1$ only standard GR and
the pseudo-complex extension of it do not contain ghost solutions, thus only
these theories are physical.
If $g^{\rm M}_{\mu\nu}$ denotes the metric of Moffat, the zero divisor components
are given by $g^{\rm M}_{\mu\nu}$ for the $\sigma_+$ component and
$\left( g^{\rm M}_{\mu\nu}\right)^T$
($T$ stands for {\it transposed})
for the $\sigma_-$ component. These components are different because Moffat
assumes a non-symmetric metric. If the metric would be symmetric, then
both components would be equal. In contrast to this, we will consider for the
two zero-divisor components of the metric
{\it two different but symmetric metrics}.
This will not change the main conclusions given in \cite{ghosts}:
In each zero-divisor component we have a standard GR with a symmetric metric
and according to \cite{ghosts} both contain only physical solutions
in the weak field limit. Afterwards we have to project to real solutions.

In pc-GR \cite{hess1} the space-time coordinates
$X^\mu$ ($\mu$ = 0,1,2,3) have pseudo-real components, which are identified
with the real space-time components $x^\mu$, while the pseudo-imaginary
components are given by $(l/c)u^\mu$, proportional to the tangent vector at a given
space-time point, which can be identified in case of a world line of a particle
as its four-velocity at a given point of the world line. The factor $l$ has the
units of a length and is
introduced for dimensional reasons. This automatically introduces a 
minimal length scale ($l$) into the theory which implies a maximal acceleration
\cite{caneloni}.
The consequences of that have not been investigated yet in a pc-GR,
but we suspect that it will be along the lines as discussed in \cite{feoli}.
The projection, as given in (\ref{projection}), then is equivalent to setting
$l=0$, or accelerated systems are excluded.

Note, that in most publications units are used where $c=1$ (as done
by us also in \cite{hess1}). Now, we opted to use $c$ with its MKS units for
practical reasons. In the MKS units, the $l/c$ has the unit of time.

In pc-GR the metric has a pseudo-real and a pseudo-imaginary component,
or the zero-divisor components,

\beqa
g_{\mu\nu} & = & g_{\mu\nu}^R + I g_{\mu\nu}^I
\nonumber \\
& = & g_{\mu\nu}^+ \sigma_+ + g_{\mu\nu}^- \sigma_-
~~~,
\label{metric}
\eeqa
are both assumed to be symmetric, i.e. we do not follow Moffat's proposal
of a non-symmetric metric, though our theory can be extended to it.
Furthermore, 

\beqa
g_{\mu\lambda} g^{\lambda\nu} & = & \delta_\mu^\nu
~~~.
\label{metric-ortho}
\eeqa
The $\sigma_\pm$ components of the metric are related to their pseudo-real
and pseudo-imaginary components via

\beqa
g_{\mu\nu}^R & = & \frac{1}{2}\left( g_{\mu\nu}^+ + g_{\mu\nu}^- \right)
\nonumber \\
g_{\mu\nu}^I & = & \frac{1}{2}\left( g_{\mu\nu}^+ - g_{\mu\nu}^- \right)
~~~.
\label{relation-g}
\eeqa
Choosing $g_{\mu\nu}^+ = g^{\rm M}_{\mu\nu}$ and
$g_{\mu\nu}^- =  \left( g^{\rm M}_{\mu\nu}\right)^T$,
leads to Moffat's 
proposal, i.e., the "-" component of the metric is the transposed
of the "+" component of the metric. This gives a connection between both
components, which determines the system. In this contribution, however,
the metric is assumed to be symmetric in both components and a connection
between them has to be defined yet.

Due to the division into pseudo-divisor components we can define a
standard GR in both sectors, but with a different symmetric metric.
Thus, parallel displacements, connections,
four derivatives, etc., are defined in {\it exactly the same way}
as in standard GR. Also, the four-derivative of the metric
is zero, permitting the definition of a universal pseudo-complex metric
(for details, see \cite{hess1}).

Due to the new variational principle, the modified Einstein equation is
given by

\beqa
{\cal R}^{\mu\nu} -\frac{1}{2} g^{\mu\nu} {\cal R} & = &
-\frac{8\pi\kappa}{c^2} T^{\mu\nu}_- \sigma_-
~~~,
\eeqa
where we have chosen the right hand side to be proportional to $\sigma_-$.
The notation on the right hand side is intentional, because the
requirement that a function is within the zero-divisor introduces in
the $\sigma_-$ component a function, which can be associated to the energy
of a yet to be determined field. As shown in \cite{hess1} this term
is responsible for the disappearance of the black hole (and the singularity), producing
an anti-gravitational effect for high central mass density, 
which halts the collapse of the mass distribution.

In \cite{hess1} a projection method was proposed, which we will briefly
explain and we shall also outline the problems related to it. The
proposed projection is performed,
requiring that the pseudo-complex length element $d\omega^2$ is real,
which leads to

\beqa
d\omega^2 & = & g_{\mu\nu} DX^\mu DX^\nu \nonumber \\
& = & g_{\mu\nu}^0 \left( dx^\mu dx^\nu +
\left( \frac{l}{c}\right)^2 du^\mu du^\nu \right)
+ 2\left( \frac{l}{c}\right)h_{\mu\nu} dx^\mu du^\nu
~~~,
\eeqa
with

\beqa
g_{\mu\nu}^0 & = & \frac{1}{2}\left( g_{\mu\nu}^+ + g_{\mu\nu}^- \right)
\nonumber \\
h_{\mu\nu}^0 & = & \frac{1}{2}\left( g_{\mu\nu}^+ - g_{\mu\nu}^- \right)
~~~.
\eeqa
Note that, the "metric" $g_{\mu\nu}^0$ is not a tensor and, therefore,
can not lower the index of $x^\mu$ nor of $u^\mu$ (see Eq. (\ref{eq10}) below).
Second, using the orthogonality
relation $g_{\mu\lambda}^\pm g^{\lambda\nu}_\pm = \delta_\mu^\nu$ we are
led to

\beqa
g_{\mu\lambda}^0 g^{\lambda\nu}_0 + h_{\mu\lambda} h^{\lambda\nu}
& = & \delta_\mu^\nu
~~~.
\label{eq10}
\eeqa
This shows that the matrix $g^{\mu\nu}_0$ is not the inverse of
$g_{\mu\nu}^0$.

Due to these properties, we propose a projection as the one applied
in the pc-field theory, mentioned above.
Once a pseudo-complex function in terms
of the coordinates and momenta is given, its projection is obtained by
substituting all pseudo-complex variables, momenta (velocities) and parameters
of the theory
by their pseudo-real parts. Also possibly appearing parameters in these
functions, like the transformation parameters $\omega^{\mu\nu}$ in a
finite Lorentz transformation, have to be substituted by their pseudo-real part.
This will be applied here too and examples will be given, like the
pc-Schwarzschild, the Reissner-Nordstr\"om and the
Kerr solution.

In general, let us denote by $G_\pm$ the geometrical space of
the $\sigma_\pm$ component. The zero-divisor metric components are given by
$g_{\mu\nu}^\pm$. The two geometrical spaces commute due to the
property of the $\sigma_\pm$ operators. This we denote via
$G_+ \otimes G_-$ which is finally reduced by the projection to the
real geometrical space $G$, i.e.

\beqa
G_+ \otimes G_- \supset G
~~~.
\eeqa
Again, the projection corresponds to set
within the metric all contributions, which are
proportional to powers in $l$, to zero. This leads to a metric, which
does not depend on the acceleration.
The case where the contributions of higher powers in
$l$ are important to the
structure of the metric in certain areas of the space-time,
i.e., the case where the metric depends on the acceleration of a system, has still to be
investigated.

For the metric, the final projection is achieved by

\beqa
g_{\mu\nu} (X, {\cal A}) \rightarrow g_{\mu\nu} (x,{\cal A}^R)
~~~,
\label{g-approx}
\eeqa
where $X$, $x$ is a shorthand notation for $(X^\lambda)$, $(x^\lambda)$
and ${\cal A}$, ${\cal A}^R$ for a set of pseudo-complex
$({\cal A}_\alpha )$ and pseudo-real parameters
$({\cal A}^R_\alpha )$ respectively.

It is interesting to know the first important contributions
of $l$ to the length element. In fact the first object is the metric itself. It
can be expanded in powers of $(l/c)u^\mu$. However, the norm of $u^\mu$ is
always smaller or equal to the speed of light.
Assuming that the minimal length scale is
of the order of the Planck length, implies that these contributions can be
safely neglected and the metric can still be approximated by (\ref{g-approx}).
However, this is not the case for $DX^\mu DX^\nu$ in the length element.
There a term $(l/c)^2 du^\mu du^\nu$ appears, i.e., it depends on the
{\it change} of the four velocity, i.e. on the acceleration. If one approaches
situations with the maximal acceleration $c^2/l$,
then the term $(l/c)^2du^\mu du^\nu$ is of the order of $l^0 = 1$ and can not be
neglected. In normal situations, where the acceleration is small,
the case considered
here, then also this term can be neglected and the length element is
of the same form as in standard GR. For large accelerations, the length
element approaches the form as used by other theories, like the one
proposed by Born \cite{born1,born2} and Caianiello \cite{caneloni}.
This is the reason why we mention it here, because, obviously these theories are a special limit of ours.
We will not further elaborate on this but refer to first attempts in this
direction \cite{feoli}.

With this motivation, for the length element squared we have

\beqa
d\omega^2 & = & g_{\mu\nu}(X,P) DX^\mu DX^\nu
\nonumber \\
& = & g_{\mu\nu}(X,P) \left( dx^\mu dx^\nu +
\left( \frac{l}{c} \right)^2 du^\mu du^\nu \right)
+ g_{\mu\nu}(X,P) 2\left( \frac{l}{c} \right)I dx^\mu du^\nu
\nonumber \\
& \rightarrow & g_{\mu\nu}(x,p) \left( dx^\mu dx^\nu +
\left( \frac{l}{c} \right)^2 du^\mu du^\nu
\right)
+ g_{\mu\nu}(x,p) 2\left( \frac{l}{c} \right)I dx^\mu du^\nu
\nonumber \\
& \rightarrow & g_{\mu\nu}(x,p) \left( dx^\mu dx^\nu +
\left( \frac{l}{c} \right)^2 du^\mu du^\nu
\right)
~~~.
\label{do2-r}
\eeqa
After mapping $X$ and $P$ to $x$ and $p$ (including the mapping of
parameters) in the third line, 
in the last step in
(\ref{do2-r}) the pseudo imaginary part
of the length square element, appearing in the second line,
has been set to zero, due to the condition that the length element has
to be pseudo-real.
It is interesting to note that the condition of disappearance of the
pseudo-imaginary part in the length element (\ref{do2-r})
leads to

\beqa
g_{\mu\nu}(x,p) dx^\mu du^\nu & = & 0
~~~,
\eeqa
which is nothing but the dispersion relation of a particle.
Normally, this dispersion relation is set to zero by hand (see, for example,
\cite{feoli}). In our procedure it is a consequence of the condition that
$d\omega^2$ is pseudo-real and it represents a subsidiary condition.

Note, that the resulting $d\omega^2$ in (\ref{do2-r}) is equal to the one used by M. Born and
related theories (setting $c=1$).

\section{The pseudo-complex Schwarzschild solution and fluid description
for the dark energy}

In this section the pc-Schwarzschild solution will be revisited.
Several steps are directly copied from \cite{adler} but reformulated
within the pseudo-complex language.

The pseudo-complex length element square for the Schwarzschild solution
has the form

\beqa
d\omega^2 & = & e^{\nu} (DX^0)^2 - e^\lambda (DR^0)^2
-R^2\left[ (D\theta )^2 + ({\rm sin}\theta )^2 (D\phi )^2 \right]
~~~,
\label{schw-dw2}
\eeqa
The Einstein equation determines the explicit structure of
$\nu$ and $\lambda$.

In \cite{hess1} the additional, arbitrary constraint ${\cal R}=0$ for the
curvature scalar was imposed. This, however, led to a unnatural large
contribution to the metric tensor, excluded by experiment.
Therefore, this condition is skipped here.

The components ${\cal R}_{\mu\nu}$ of the Ricci tensor in the
Einstein equation reduce to ${\cal R}_{\mu\nu} \epsilon {\cal P}^0$.
${\cal R}_{\mu\nu}$ is the pc-Ricci tensor. As defined in \cite{hess1},
we have

\beqa
{\cal R}_{00} & = & -\frac{1}{2}e^{\nu-\lambda}\xi_0 \sigma_-
\nonumber \\
{\cal R}_{11} & = & \frac{1}{2}\xi_1 \sigma_-
\nonumber \\
{\cal R}_{22} & = & \xi_2 \sigma_-
\nonumber \\{\cal R}_{33} & = & \xi_3 \sigma_-
~=~ \xi_2 {\rm sin}^2\theta \sigma_-
~~~.
\label{R}
\eeqa
Denoting with a prime the derivation with respect to $R$,
the former condition $\left( \nu^\prime + \lambda^\prime \right)=0$ of the
standard GR changes now to

\beqa
\left( \nu^\prime + \lambda^\prime \right) & = & \frac{1}{2}R
\left( \xi_0 - \xi_1 \right) \sigma_-
~~~.
\label{xi0-xi1}
\eeqa
In \cite{hess1} it was assumed for simplicity that $\xi_0 = \xi_1$. This,
however, led together with ${\cal R}=0$
to corrections in the metric of order $\frac{1}{R^2}$,
which are excluded by the experiment \cite{exp-ppn}.
{\it This led us to reconsider the pc-Schwarzschild solution, skipping the
above conditions}.

We rewrite the Einstein equation into

\beqa
{\cal R}_\mu^\nu - \frac{1}{2}g_\mu^\nu {\cal R} & = & \Xi_\mu^\nu \sigma_-
~~~.
\label{Eeqspsym}
\eeqa
where we used the abbreviation

\beqa
\Xi_\mu^\nu & = & -\frac{8\pi\kappa}{c^2} T_\mu^\nu
~~~,
\label{XiT}
\eeqa
with $T_\mu^\nu$ representing the components of the energy-momentum tensor.
We also rewrote the equation into upper and lower index notation.

Because the Ricci tensor is diagonal, the following definition is also used

\beqa
\Xi_\mu^\mu & = & \Xi_\mu
\label{XiXi}
\eeqa
and from that we obtain a relation between $\Xi_\mu$ and $\xi_\mu$, i.e.,

\beqa
-\frac{1}{4}e^{-\lambda_-} \xi_0+\frac{1}{4}e^{-\lambda_-}\xi_1+\frac{\xi_2}{R_-^2}
& = & \Xi_0
\nonumber \\
\frac{1}{4}e^{-\lambda_-} \xi_0-\frac{1}{4}e^{-\lambda_-}\xi_1+\frac{\xi_2}{R_-^2}
& = & \Xi_1
\nonumber \\
\frac{1}{4}e^{-\lambda_-}\left( \xi_0 + \xi_1 \right) & = & \Xi_2
\nonumber \\
\Xi_3 & = & \Xi_2
~~~.
\label{gl-Xi}
\eeqa

In principle,
we can abolish the use of the $\xi_\mu$ functions and keep only the
$\Xi_\mu$. However, because we introduced the $\xi_\mu$ functions
in \cite{hess1} and want to compare with the results there, we will
keep the $\xi_\mu$ functions and express $\Xi_\mu$ in terms of $\xi_\mu$.

The equations in (\ref{gl-Xi}) are resolved for $\xi_\mu$, leading us to

\beqa
\frac{2\xi_2}{R_-^2} & = & \Xi_0+\Xi_1
\nonumber \\
\frac{1}{2}e^{-\lambda_-}\left( \xi_0-\xi_1 \right) & = & \Xi_1-\Xi_0
\nonumber \\
\frac{1}{4}e^{-\lambda_-}\left( \xi_0 + \xi_1 \right) & = & \Xi_2
~~~.
\label{gl-3}
\eeqa
Multiplying in a first step the last equation with 2 and adding to it
the second equation and subtracting in a second step the second equation
from the last one, we obtain

\beqa
e^{-\lambda_-} \xi_0 & = & 2\Xi_2 + \Xi_1 - \Xi_0 \nonumber \\
e^{-\lambda_-} \xi_1 & = & 2\Xi_2 - \Xi_1 + \Xi_0
\nonumber \\
\frac{2\xi_2}{R^2} & = & \Xi_0 + \Xi_1
~~~.
\label{eq-3}
\eeqa
Obviously the last equation in (\ref{eq-3}) is a repetition
of the first equation in (\ref{gl-3}).

Now, we use for $T_\mu^\nu = \text{diag}(\rho, -\frac{p}{c^2}, -\frac{p}{c^2}, -\frac{p}{c^2})$ the expression for an ideal fluid/gas \cite{adler} 
together with (\ref{XiT}), which gives

\beqa
\Xi_0 & = & -\frac{8\pi \kappa}{c^2} \rho
\nonumber \\
\Xi_k & = & \frac{8\pi \kappa}{c^2} \frac{p}{c^2} ~~~,~~~ (k=1,2,3)
~~~. \label{Xiidealfluid}
\eeqa

Substituting this into equation (\ref{eq-3}) gives

\beqa
\frac{2\xi_2}{R_-^2} & = &
-\frac{8\pi\kappa}{c^2} \left( \rho - \frac{p}{c^2} \right)
\nonumber \\
e^{-\lambda_-} \xi_0 & = &
\frac{8\pi\kappa}{c^2} \left( 3\frac{p}{c^2}+ \rho \right)
\nonumber \\
e^{-\lambda_-} \xi_1 & = &
-\frac{8\pi\kappa}{c^2} \left( \rho - \frac{p}{c^2} \right)
~~~,
\label{xik-rho}
\eeqa
which immediately tells us that

\beqa
e^{-\lambda_-} \xi_1 & = & \frac{2\xi_2}{R_-^2}
~~~.
\label{xi1-xi2}
\eeqa
This result is independent of the assumption that $\xi_0=\xi_1$!
(\ref{xi1-xi2}) can also be obtained by noting that $\Xi_1 = \Xi_2$ (see (\ref{Xiidealfluid}))
and using (\ref{gl-Xi}).

\subsection{Solving the Einstein equation}

We will concentrate on the $\sigma_-$ component, because the
$\sigma_+$ component is the same as in \cite{adler}.

A repeated differentiation of (\ref{xi0-xi1}) leads to 

\beqa
\nu_-^{\prime\prime} & = & -\lambda_-^{\prime\prime}
+ \frac{1}{2}\left( \xi_0 - \xi_1 \right)
+ \frac{R_-}{2}\left( \xi_0^\prime - \xi_1^\prime \right)
~~~.
\label{nu-pp}
\eeqa

Copying the steps of \cite{hess1}, Eqs. (52)
\beqa
{\cal R}_{00} & = &
-\frac{e^{\nu-\lambda}}{2} \left( \nu^{\prime\prime}
+\frac{\nu^{\prime 2}}{2} - \frac{\lambda^\prime \nu^\prime}{2}
+\frac{2\nu^\prime}{R} \right) \nonumber \\
{\cal R}_{11} & = &
\frac{1}{2} \left( \nu^{\prime\prime}
+\frac{\nu^{\prime 2}}{2} - \frac{\lambda^\prime \nu^\prime}{2}
-\frac{2\lambda^\prime}{R} \right) \nonumber \\
{\cal R}_{22} & = &
\left( e^{-\lambda} R_- \right)^\prime - 1 \nonumber \\
{\cal R}_{33} & = &
\sin^2\theta ~ \left[ \left( e^{-\lambda} R_- \right)^\prime - 1 \right] ~~~.
\label{rmn}
\eeqa
and (53)
\beqa
\nu^{\prime\prime} + \frac{1}{2} \nu^{\prime  2} - \frac{1}{2} \lambda^\prime \nu^\prime
+ \frac{2\nu^\prime}{R_-} & = & \xi_0 \sigma_-
\nonumber \\
\nu^{\prime\prime} + \frac{1}{2} \nu^{\prime 2} - \frac{1}{2} \lambda^\prime \nu^\prime
- \frac{2\lambda^\prime}{R_-} & = & \xi_1 \sigma_-  ~~~.
\label{r00r11}
\eeqa
 and restricting to the
$\sigma_-$ component only, we obtain from
the equation ${\cal R}^-_{11} = \frac{1}{2}\xi_1$ the following one

\beqa
\nu_-^{\prime\prime} + \frac{1}{2}\nu_-^{\prime 2}
- \frac{1}{2}\lambda_-^\prime \nu_-^\prime - \frac{2\lambda_-^\prime}{R}
& = & \xi_1
~~~.
\label{nupp}
\eeqa
This is the same as in \cite{hess1}. Because in general $\xi_0 \neq \xi_1$,
in the equation $R^-_{22}=\xi_2$ (see (\ref{R})) an additional term appears
in $R^-_{22}$, namely $R_- e^{-\lambda_-} \left( \frac{\nu_-^\prime + \lambda_-^\prime}
{2} \right)$, which is proportional to $\left( \lambda_-^\prime + \nu_-^\prime \right)$,
not equal to zero. Its origin is shown in \cite{adler}, Eq. (6.44).
Taking this term into account changes the equation $R^-_{22}=\xi_2$ to

\beqa
\left[ R_- e^{-\lambda_-}\right]^\prime -1 +
R_- e^{-\lambda_-} \left( \frac{\nu_-^\prime + \lambda_-^\prime}
{2} \right) & = & \xi_2
~~~.
\label{r22}
\eeqa
Using (\ref{xi0-xi1}), which relates $\nu_-^\prime$ to
$\lambda_-^\prime$ and $\xi_0$, $\xi_1$, the last equation
converts to

\beqa
e^{-\lambda_-} \left[ 1+\frac{R_-\nu_-^\prime}{2}-\frac{R_-\lambda_-^\prime}{2}
\right] -1 & = & \xi_2
~~~.
\label{corr}
\eeqa
When $\xi_0 = \xi_1$, then $\nu_-^\prime = - \lambda_-^\prime$ and the old
result of \cite{hess1} is obtained.

Substituting again (\ref{xi0-xi1}) and its derivative, given by
(\ref{nu-pp}),
into the left hand side of (\ref{nupp}) and reordering terms leads to

\beqa
& -\left(\lambda_-^{\prime\prime} - \lambda_-^{\prime 2}
+\frac{2\lambda_-^\prime}{R_-} \right)&
\nonumber \\
& +\frac{1}{2}\left(\xi_0-\xi_1\right)
+\frac{1}{2}R_-\left(\xi_0^\prime - \xi_1^\prime \right)
-\frac{3}{4}\lambda_-^\prime R_- \left(\xi_0-\xi_1\right)
+\frac{1}{8} R_-^2 \left(\xi_0-\xi_1\right)^2 &
\nonumber \\
& = \xi_1 &
~~~.
\label{casi}
\eeqa
In the first line we have
$-\left(\lambda_-^{\prime\prime} - \lambda_-^{\prime 2}
+\frac{2\lambda_-^\prime}{R_-} \right)$ =
$\frac{e^{\lambda_-}}{R_-}\left( R_-e^{-\lambda_-}\right)^{\prime\prime}$,
which can be directly verified by executing the second derivative. 
Shifting $\left[ R_- e^{-\lambda_-} \right]'$ in
(\ref{r22}) to one side and substituting this into
$\frac{e^{\lambda_-}}{R_-}\left( R_-e^{-\lambda_-}\right)^{\prime\prime}$,
we arrive at

\begin{align}
-\left(\lambda_-^{\prime\prime} - \lambda_-^{\prime 2}
+\frac{2\lambda_-^\prime}{R_-} \right)& = \frac{e^{\lambda_-}}{R_-}\left( R_-e^{-\lambda_-}\right)^{\prime\prime} \notag \\
& = \frac{e^{\lambda_-}}{R_-}\left( 1+\xi_2
- \frac{R_-^2}{4}\left( \xi_0-\xi_1\right) e^{-\lambda_-}\right)^\prime \notag \\
&= \frac{e^{\lambda_-}}{R_-}  \xi_2' - \frac{1}{2}\left( \xi_0-\xi_1\right) - \frac{R_-}{4} \left( \xi_0'-\xi_1'\right) + \frac{\lambda_- R_-}{4}\left( \xi_0-\xi_1\right)
~~~.
\end{align}
This can be inserted in the first line in (\ref{casi}) to obtain
\begin{align}
\frac{e^{\lambda_-}}{R_-}  \xi_2'   
+\frac{1}{4}R_-\left(\xi_0^\prime - \xi_1^\prime \right)
-\frac{1}{2}\lambda_-^\prime R_- \left(\xi_0-\xi_1\right)
+\frac{1}{8} R_-^2 \left(\xi_0-\xi_1\right)^2  = \xi_1 \quad . \label{eq:casi2}
\end{align}
 By also
expressing the $\xi_1$ on the right hand side in (\ref{eq:casi2}) by
$\xi_2$ using (\ref{xi1-xi2}),
we obtain finally

\beqa
& \frac{e^{\lambda_-}}{R_-}\xi_2^\prime - e^{\lambda_-}\frac{2\xi_2}{R_-^2} &
\nonumber \\
& = &
\nonumber \\
& -\frac{1}{4}R_-\left( \xi_0^\prime-\xi_1^\prime\right)
+\frac{1}{2}\lambda_-^\prime R_- \left( \xi_0-\xi_1\right)
-\frac{1}{8}R_-^2 \left( \xi_0-\xi_1\right)^2 &
~~~.
\label{difa}
\eeqa
This is a differential equation relating $\xi_2$ with $\xi_0$.

Using the derivative of $R_-e^{-\lambda_-}$ as given in (\ref{r22}),
integrating it
and setting the integration constant equal to $-2{\cal M}_-$,
we obtain

\beqa
R_-e^{-\lambda_-} & = & R_- - 2{\cal M}_- + \int \xi_2 dR_-
- \frac{1}{4} \int e^{-\lambda_-}R_-^2 \left(\xi_0-\xi_1\right)dR_-
~~~.
\nonumber \\
\label{g11-1}
\eeqa
The sum of the terms within the integral is nothing but
$\Xi_0$, as given in (\ref{gl-Xi}).

Now, we can use the connection of the $\xi_\mu$ functions in terms of
the density $\rho$ and pressure $p$, given in (\ref{xik-rho}), 
with the result

\beqa
R_- e^{-\lambda_-} & = & R_- - 2{\cal M}_-
-\frac{8\pi\kappa}{c^2}\int R_-^2 \rho dR_-
~~~.
\label{g11}
\eeqa
{\it The same we would get by setting $\xi_0 = \xi_1$, i.e., $e^{-\lambda_-}$
depends only on the density}.

However, $e^{-\nu_-}$, the metric component of
$(DX_-^0)^2$, will not have such a simple expression. It is given by

\beqa
e^{\nu_-} & = & e^{-\lambda_-}
e^{\frac{1}{2} \int R_-\left(\xi_0-\xi_1\right)dR_-}
\nonumber \\
e^{-\lambda_-}
e^{\frac{8\pi\kappa}{c^2} \int R_- e^{\lambda_-}
\left[ \left(\frac{p}{c^2}\right) + \rho \right]dR_-} & = &
e^{-\lambda_-}e^{f_-}
~~~.
\label{e-g}
\eeqa
The factor $e^{\frac{1}{2} \int R_-\left(\xi_0-\xi_1\right)dR_-}$
is a positive function in $R_-$.

We now proceed in rewriting the expression in (\ref{g11}), considering first
the last term:

\beqa
-\frac{8\pi\kappa}{c^2}\int R_-^2 \rho dR_- & = &
\frac{2\kappa}{c^2} M_{\rm de}(R_-)
~~~.
\label{add}
\eeqa
$M_{\rm de}(R_-)$ is the accumulated mass of the dark energy. 
The density will be proportional to a negative power of $R_-$, representing a
decline of the dark energy density with increasing distance. Thus, the
integral in $R_-$ of the dark energy density will have a negative sign. 
This is accounted for on
the left hand side of (\ref{add}) by the multiplication with minus one,
rendering the dark energy mass positive.

The metric element $g_-^{11}$ is equal to $e^{-\lambda_-}$, which
can be obtained from (\ref{g11}) and substituting into this equation
the integral by (\ref{add}). 
The expression for $g_-^{11}$ is finally given by

\beqa
e^{-\lambda_-} & = & 1-\frac{2{\cal M}_-}{R_-}
+\frac{1}{R_-}\frac{2\kappa}{c^2}M_{\rm de}(R_-)
\nonumber \\
& = & 1-\frac{2{\cal M}_-}{R_-}+\frac{2m_{de}(R_-)}{R_-}
~~~,
\nonumber \\
m_{de} & = & \frac{\kappa M_{\rm de}(R_-)}{c^2}
~~~.
\label{e-lambda}
\eeqa
In (\ref{e-g}) the $g^-_{00} = e^{\nu_-}$ was related to
$e^{-\lambda_-}$, which is the $g^{11}_-$ metric component.
An additional factor appeared, abbreviated by  $e^{f_-}$.
With this and (\ref{e-lambda}),
the $g_{00}^-$ component acquires the structure

\beqa
g_{00}^- & = & \left( 1- \frac{2{\cal M_-}}{R_-}+ \frac{2m_{{\rm de}}}{R_-}
\right) e^{f_-}
~~~,
\label{g00}
\eeqa
where the function $f_-$ has yet to be determined.  In \cite{hess1}
the condition $g_{00}>0$ {\it after projection} was imposed and the
consequences are similar with respect to the redshift, the difference
lying in the extra factor $e^f$, which is due to the assumption
that in general $\xi_0 \neq \xi_1$.

Having resolved the components of the metric tensor, we will proceed
to determine a relation between the pressure and the density of the
energy outside the central mass.

In (\ref{xik-rho}) the $\xi_\mu$ functions were related to the
pressure, $p$, and the density, $\rho$. Substituting these relations
into the differential equation (\ref{difa}), after a short
rearrangement we arrive at

\beqa
\frac{p^\prime}{c^2} & = & \frac{1}{2}\lambda^\prime
\left( \frac{p}{c^2} + \rho \right)
-\frac{8\pi\kappa}{c^2}\frac{1}{2}R_- e^{\lambda_-}
\left( \frac{p}{c^2} + \rho \right)^2
~~~.
\label{difb}
\eeqa
Obviously the derivative with respect to $\rho$ does not appear!

Deriving $e^{-\lambda_-}$ in (\ref{e-lambda}) with respect to $R_-$
yields

\beqa
-\lambda_-^\prime e^{-\lambda_-} & = & \frac{2}{R_-^2}
\left( {\cal M}_- - R_- m_{{\rm de}}(R_-) + R_- m_{{\rm de}}^\prime (R_-)
\right)
\eeqa
and we arrive at

\beqa
p & = & p(\rho )
\nonumber \\
m_{{\rm de}}^\prime & = & - \frac{4\pi\kappa}{c^2} R_-^2 \rho
\nonumber \\
\frac{p^\prime}{c^2} & = &
-\frac{
\left( \frac{4\pi\kappa}{c^2} \frac{p}{c^2} R_-^3
+ {\cal M}_- - m_{{\rm de}}(R_-) \right)
}{
R\left(R-2{\cal M}_- + 2m_{{\rm de}}(R_-) \right)
}
\left( \frac{p}{c^2} + \rho \right)
\nonumber \\
e^{\lambda_-} & = & \left( 1- \frac{2\cal M}{R_-}
+ \frac{2m_{{\rm de}}(R_-)}{R_-} \right)
\nonumber \\
\nu^\prime & = &
2\frac{\left( \frac{4\pi\kappa}{c^2}\frac{p}{c^2} R_-^3 +{\cal M}_-
- m_{{\rm de}}(R_-)\right)
}{
R_-\left( R_--2{\cal M}_- + 2 m_{{\rm de}}(R_-)\right)
}
~~~,
\label{sum-eq}
\eeqa
where we have added the equations for $\nu (R_-)$, $e^{\lambda_-}$, the
relation of $\rho$ to the derivative of $m_{{\rm de}}$ with respect to $R_-$
and a {\it still unknown} equation of state $p=p(\rho )$. This set of
equations is equivalent to the one of (14.25) in \cite{adler}, which were
obtained within a model for a relativistic star structure. The difference
is in the substitution of $m(r)$ by
$\left( {\cal M}_- - m_{{\rm de}}(R_-) \right)$.
Note, that when $m_{{\rm de}}$ increases, for an outside observer
the effective mass of the object decreases.

\subsection{The pc-Schwarzschild metric}
\label{pcSchw}

In the last sub-section we obtained an analytic solution for the $\sigma_-$
component of the metric. The one for the $\sigma_+$ component is
identical to the one derived by e.g. Adler et al. \cite{adler}.
In the $\sigma_-$ component the functions $\Omega_-$ and
$f_-$ appear, the last in the $g^-_{00}$ metric component.
In the $\sigma_+$ component, however, no such functions appear. In order
to rewrite the $\sigma_+$ component in a form similar to the one in the
$\sigma_-$ component, we introduce the definitions
$\Omega_+ = 0$ and $f_+=0$. With the help of this both components
can be written as

\beqa
\left( g_{\mu\nu}^\pm \right) & = &
\left(
\begin{array}{cccc}
\left( 1 - \frac{2{\cal M}_\pm}{R_\pm} + \frac{\Omega_\pm}{R_\pm} \right)
e^{f_\pm}
& 0 & 0 & 0 \\
0 & -\left( 1 - \frac{2{\cal M}_\pm}{R_\pm} + \frac{\Omega_\pm}{R_\pm}
\right)^{-1}
& 0 & 0 \\
0 & 0 & -R_\pm^2 & 0 \\
0 & 0 & 0 & -R_\pm^2 {\rm sin}^2\theta
\end{array}
\right)
\nonumber \\
~~~.
\eeqa
As can be seen, the metric tensor has the equivalent functional form
in both the $\sigma_-$ and $\sigma_+$ component.
We have used the notation

\beqa
R_\pm & = & r \pm lI{\dot r} \nonumber \\
{\cal M} & = & {\cal M}_+ \sigma_+ + {\cal M}_- \sigma_-
\nonumber \\
{\cal M}_\pm & = & m
\nonumber \\
\Omega & = & 2m_{{\rm de}}\sigma_- ~=~ \Omega_+ \sigma_+ + \Omega_- \sigma_-
\nonumber \\
\Omega_+ & = & 0 ~~~,~~~ \Omega_- ~=~ 2m_{{\rm de}}
\nonumber \\
f_+ & = & 0 ~~~,~~~ f_-~\neq~0
~~~.
\eeqa
The condition ${\cal M}_\pm = m$ comes from the requirement that the
standard GR should be contained in the limit of large distances $r$,
which will be verified further below.
The pseudo-real elements of the parameters are

\beqa
{\ M}_R = m ~,~
\Omega_R = \frac{1}{2}\left( \Omega_+ + \Omega_- \right) ~=~
\frac{\Omega_-}{2} ~=~ m_{{\rm de}}
~~~.
\eeqa

Because now the metric tensors in both $\sigma$-component have the same
functional form,
the total pseudo-complex metric can be written as
$g_{\mu\nu}(\Omega , f, R)$ =
$g^+_{\mu\nu}(\Omega_+ , f_+, R_+)\sigma_+$ +
$g^-_{\mu\nu}(\Omega_- , f_-, R_-)\sigma_-$,
which gives

\beqa
\left( g_{\mu\nu} \right) & = &
\left(
\begin{array}{cccc}
\left( 1 - \frac{2{\cal M}}{R} + \frac{\Omega}{R} \right) e^f
& 0 & 0 & 0 \\
0 & -\left( 1 - \frac{2{\cal M}}{R} + \frac{\Omega}{R}
\right)^{-1}
& 0 & 0 \\
0 & 0 & -R^2 & 0 \\
0 & 0 & 0 & -R^2 {\rm sin}^2\theta
\end{array}
\right)
\nonumber \\
\eeqa
and the projected metric, following our prescription, is

\beqa
\left( g_{\mu\nu}(r) \right) & = &
\left(
\begin{array}{cccc}
\left( 1 - \frac{2m}{r} + \frac{\Omega_-}{2r} \right)
e^{\frac{f_-}{2}}
& 0 & 0 & 0 \\
0 & -\left( 1 - \frac{2m}{r} + \frac{\Omega_-}{2r}
\right)^{-1}
& 0 & 0 \\
0 & 0 & -r^2 & 0 \\
0 & 0 & 0 & -r^2 {\rm sin}^2\theta
\end{array}
\right)
~~~.
\nonumber \\
\eeqa

The length squared element is, therefore, given by (see (\ref{do2-r}))

\beqa
d\omega^2 & = & \left( 1 - \frac{2m}{r} + \frac{\Omega_-}{2r} \right)
e^{\frac{f_-}{2}}
(dx^0)^2 - \left( 1 - \frac{2m}{r} + \frac{\Omega_-}{2r} \right)^{-1}
(dr)^2 - r^2 \left( (d\theta )^2 + {\rm sin}^2\theta (d\phi )^2 \right)
~~~.
\nonumber \\
\eeqa
In \cite{hess1} we imposed the condition $g_{00}(r)>0$, so that the
signature for the time stays the same. 
In Chapter 5 and 6 we will consider a corretion of $\Omega_- = \frac{B}{r^2}$ (with $B>0$) as a model. The condition $g_{00}(r) > 0$ then is
\begin{equation}
        1 - \frac{2m}{r} + \frac{B}{2r^3} > 0 \quad . \label{eq:g00groesser0}
\end{equation}
To find a value for $B$ which satisfies this condition for all $r > 0$ we will have a look at the extremal value of $g_{00}$. As we know from the limiting behaviour of $g_{00}$ ($g_{00} \rightarrow 1 $ for $r \rightarrow \infty$ and $g_{00} \rightarrow + \infty$ for $r \rightarrow 0$) its extremal value will be a minimum. A quick calculation gives $r = \left(\frac{3}{4} \frac{B}{m}\right)^{1/2}$ as the value of the minimum of $g_{00}$. Inserting this in (\ref{eq:g00groesser0}) yields
\begin{equation}
        \left(\frac{3}{4} \frac{B}{m}\right)^{3/2} - B > 0 \quad \Rightarrow \quad  B > \frac{64}{27} m^3 \quad .
 \label{eq:bedingungfuerB}
\end{equation}

The conclusions in \cite{hess1}
for the redshift in a Schwarzschild solution remain the same, i.e., after
an increase of the redshift a minimum is reached. Because the potential is
proportional to the square root of $g_{00}(r)$\cite{misner}, this indicates a minimum
in the potential, which is repulsive for lower radial distances $r$.
As a consequence, the collapse of a star is halted latest at the minimum
and it can not contract to a singularity. The star probably still oscillates
around this minimum, which should be eventually observable.
Further investigations concerning this aspect are required. 

\section{The pseudo-complex Reissner-Nord\-str\"om solution}

In this section we present our findings concerning the pc-Reissner-Nordstr\"om
solution. Details can be found in  \cite{caspar}.

Since we consider a central, charged mass at rest, the spherical
symmetry is conserved and the line element squared is again given
by (\ref{schw-dw2}). Furthermore, we can adopt the Einstein equation
(\ref{Eeqspsym}) after adding the energy-momentum tensor for the
electromagnetic field and the related conditions (\ref{XiT}),
(\ref{XiXi}) remain valid. Hence the Einstein equation is

\beqa
{\cal R}_\mu^\nu - \frac{1}{2}g_\mu^\nu {\cal R} & = & \Xi_{\mu}^{\nu~RN} \sigma_- - \frac{8\pi\kappa}{c^2} T_\mu^{\nu ~em}
~~~, \label{eq:einsteinReissner}
\eeqa
where $T_\mu^{\nu~ em}$ is given by \cite{adler}
  
\beqa
T_\mu^{\nu ~em} = \frac{\epsilon^2}{2c^2r^4} 
\left (
\begin{array}{cccc}
1 & 0 & 0 & 0 \\
0 & 1 & 0 & 0 \\
0 & 0 & -1 & 0 \\
0 & 0 & 0 & -1
\end{array}
\right )
~~~. 
\nonumber \\ \label{Tmunuem}
\eeqa
Thereby $\epsilon$ depends on the charge Q in the following way 
\beqa
\epsilon = \frac{Q}{4\pi\epsilon_0} \label{echarge}
~~~.
\eeqa
 
It is not certain, whether $\Xi_\mu^{\nu RN}$ is the same as in the Schwarzschild
case, because there might be a coupling of the ``dark'' energy with the
charge of the central mass. However at the moment we consider it as
improbable that $\Xi_\mu^{\nu RN}$ is equal to $\Xi_\mu^{\nu}$, since
the combination with the ideal fluid ansatz leads to an unphysical result (for detailed calculations the reader is referred to the appendix). Therefore, so far, no plausible approach exists for connecting the Reissner-Nord\-str\"om source with the Schwarzschild one.
    
Nevertheless $\xi_\mu^{{\rm RN}}$ can be defined such that a relation similar to (\ref{gl-Xi}) keeps valid, i.e.

\beqa
-\frac{1}{4}e^{-\lambda_{RN-}} \xi^{RN}_0+\frac{1}{4}e^{-\lambda_{RN-}}\xi^{RN}_1+\frac{\xi^{RN}_2}{R_-^2}
& = & \Xi^{RN}_0
\nonumber \\
\frac{1}{4}e^{-\lambda_{RN-}} \xi_0-\frac{1}{4}e^{-\lambda_{RN-}}\xi_1+\frac{\xi^{RN}_2}{R_-^2}
& = & \Xi^{RN}_1
\nonumber \\
\frac{1}{4}e^{-\lambda_{RN-}}\left( \xi^{RN}_0 + \xi^{RN}_1 \right) & = & \Xi^{RN}_2
~~~.
\eeqa

\subsection{Solving the Einstein equation} \label{RNSEq}

As in the previous chapter the $\sigma_+$ component does not differ
from the usual GR field equations, which can be obtained
in the same way as done in \cite{adler}. Thus we only have to solve for
the $\sigma_-$ component. In other words we are justified
to restrict to the $\sigma_-$ component only.
We take into account, that the energy-momentum tensor of the electromagnetic
contribution, $T_\mu^{\nu ~em}$, is real, i.e.,
$T_\mu^{\nu ~em}$ = $T_\mu^{\nu ~em}\left( \sigma_- + \sigma_+ \right)$. Thus
the $\sigma_+$ component is the same as the $\sigma_-$ component. We use
that $\left( \sigma_- + \sigma_+ \right)=1$. In addition we observe that
\begin{equation}
 {\cal R} =  {\cal R}_{0}^0 + {\cal R}_{1}^1 + {\cal R}_{2}^2 + {\cal R}_{3}^3 \quad .
\end{equation}

With this and $g^\nu_\mu = \delta^\nu_\mu$, the Einstein equations (\ref{eq:einsteinReissner}) for the $\sigma_-$ component are

\begin{align}
\frac 12 \left ( {\cal R}_{-0}^0 - {\cal R}_{-1}^1 - {\cal R}_{-2}^2 - {\cal R}_{-3}^3 \right ) &= \Xi^{RN}_0 - \frac{8\pi \kappa}{c^2} T_0^{0 ~em} \nonumber \\
\frac 12 \left ( {\cal R}_{-1}^1 - {\cal R}_{-0}^0 - {\cal R}_{-2}^2 - {\cal R}_{-3}^3 \right ) &= \Xi^{RN}_1 - \frac{8\pi \kappa}{c^2} T_1^{1 ~em}\nonumber \\
\frac 12 \left ( {\cal R}_{-2}^2 - {\cal R}_{-0}^0 - {\cal R}_{-1}^1 - {\cal R}_{-3}^3 \right ) &= \Xi^{RN}_2 - \frac{8\pi \kappa}{c^2} T_2^{2 ~em} \nonumber \\
\frac 12 \left ( {\cal R}_{-3}^3 - {\cal R}_{-0}^0 - {\cal R}_{-1}^1 - {\cal R}_{-2}^2 \right ) &= \Xi^{RN}_3 - \frac{8\pi \kappa}{c^2} T_3^{3 ~em} \label{RNEinst}
~~~.
\end{align}

Taking the difference between the second and the third
equation yields

\beqa
{\cal R}_{-2}^2 - {\cal R}_{-3}^3 = \Xi^{RN}_2 - \Xi^{RN}_3 - \frac{8\pi \kappa}{c^2} \left ( T_2^{2 ~em} - T_3^{3 ~em} \right )
\eeqa
and since the spherical symmetry and equation (\ref{Tmunuem}) demand\\
${\cal R}_{-2}^2 - {\cal R}_{-3}^3 =
\frac{8\pi \kappa}{c^2} \left ( T_2^{2 ~em} - T_3^{3 ~em} \right )= 0$
, we obtain (note, that the lower index "-" refers to the
$\sigma_-$ component and {\it not} to the sign of a number) 

\beqa
\Xi^{RN}_2 = \Xi^{RN}_3 
~~~. 
\eeqa

The difference of the zeroth and first equation leads to 

\beqa
\lambda_{RN-}^\prime + \nu_{RN-}^\prime = R_- e^{\lambda_{RN-}} \left ( \Xi^{RN}_1 - \Xi^{RN}_0 \right ) = \frac{R_-}{2} \left ( \xi^{RN}_0 - \xi^{RN}_1 \right ) \label{RNxi0-xi1}
~~~.
\eeqa
After differentiation we get

\beqa
\nu_{RN-}^{\prime\prime} &= -\lambda_{RN-}^{\prime\prime} + \frac{1}{2} \left (\xi^{RN}_0 - \xi^{RN}_1 \right ) + \frac {R_-}{2} (\xi^{RN\prime}_0 - \xi^{RN\prime}_1) \label{RNnu-pp}
~~~,  
\eeqa
which is similar to equation (\ref{nu-pp}).

By adding two times the second equation of (\ref{RNEinst}) to the
last difference and multiplying with 2e$^{\lambda_{RN-}}$ we obtain

\beqa
\nu_{RN-}^{\prime\prime} - \frac{\lambda_{RN-}^\prime \nu_{RN-}^\prime}{2} + \frac{\nu_{RN-}^{\prime 2}}{2} - \frac{2\lambda_{RN-}^\prime}{R_-} = \xi^{RN}_1 - \frac{2A}{R_-^4}e^{\lambda_{RN-}} \label{RNdeq}
~~~,
\eeqa
where we used the abbreviation (see (\ref{echarge}) for the definition of $\epsilon$)

\beqa
A := - \frac{4\pi \kappa \epsilon^2}{c^4} \label{defA}
~~~.
\eeqa

Moreover, after including (\ref{RNxi0-xi1}) the sum of the
equations with the index 0 and 1 in (\ref{RNEinst}) leads to

\beqa
\left (R_- e^{-\lambda_{RN-}}\right )' & =  & 1 + \xi^{RN}_2 - \frac 14 R_-^2 e^{-\lambda_{RN-}} \left ( \xi^{RN}_0 - \xi^{RN}_1 \right ) + \frac{A}{R_-^2} 
~~~, \nn\\
\label{RNg11dif}
\eeqa
which gives after an integration

\begin{align}
e^{-\lambda_{RN-}}  = &  1 - \frac{2M_-}{R_-} + \frac{1}{R_-} \int \xi^{RN}_2 dR_- \notag \\
& + \frac{1}{4R_-} \int e^{-\lambda_{RN-}} R_-^2 \left ( \xi^{RN}_0 - \xi^{RN}_1 \right ) dR_- - \frac{A}{R_-^2} ~~~. \nn\\
\label{RNg11}
\end{align}

Now we can generate the Reissner Nordstr\"om equivalent of (\ref{difa}) through combining (\ref{RNxi0-xi1}), (\ref{RNnu-pp}) and (\ref{RNg11dif}) with (\ref{RNdeq}) (a detailed calculation can be found in appendix (\ref{eq:lambdaRN}) )

\beqa
& \frac{e^{\lambda_{RN-}}}{R_-}\xi_2^{RN\prime} - \xi^{RN}_1 &
\nonumber \\
& = &
\nonumber \\
& -\frac{1}{4}R_-\left( \xi^{RN\prime}_0-\xi^{RN\prime}_1 \right)
+\frac{1}{2}\lambda^{RN\prime}_- R_- \left( \xi^{RN}_0-\xi^{RN}_1\right)
-\frac{1}{8}R_-^2 \left( \xi^{RN}_0-\xi^{RN}_1\right)^2 &
~~~.
\nonumber \\ \label{diffeqRN}
\eeqa  

This differential equation relates $\xi^{RN}_0$, $\xi^{RN}_1$ and
$\xi^{RN}_2$, whereby $\xi^{RN}_1$ can again be expressed through $\xi^{RN}_2$,
if the ideal fluid ansatz is used.

Through the use of (\ref{RNxi0-xi1}) and (\ref{RNg11}) we can
calculate the $g^-_{00}$-component of the metric

\beqa
g^-_{00} & =  & e^{\nu_{RN-}} = e^{-\lambda_{RN-}} e^{\frac{1}{2} \int R_- \left ( \xi^{RN}_0 - \xi^{RN}_1 \right ) dR_-} 
~~~.
\eeqa

Within the ideal fluid ansatz the metric terms can be written in a form similar as in chapter 3

\beqa
g^-_{11} & = & \left ( 1 - \frac{2M_-}{R_-} + \frac{2m^{RN}_{de}(R_-)}{R_-} - \frac{A}{R_-^2} \right )^{-1} \\
g^-_{00} & = & \left ( 1 - \frac{2M_-}{R_-} + \frac{2m^{RN}_{de}(R_-)}{R_-} - \frac{A}{R_-^2} \right ) e^{f_{A-}}
~~~, 
\eeqa
where, equivalent to (\ref{g00}), the function $f_{A-} = \frac{1}{2} \int R_- \left ( \xi^{RN}_0 - \xi^{RN}_1 \right ) dR_- $ has still to be determined. 

\subsection{The pc-Reissner-Nordstr\"om solution}
After these calculations within the last sub-section \ref{RNSEq} we now know the analytic
solution of both components for the Reissner-Nordstr\"om problem.
Hence we are now able to determine the real metric. We use the same
notations as in \ref{pcSchw} and additionally define $f_{A+}$ = 0
(remember that we introduced this definition in order to have the same
functional form of the metric ternsor in both $\sigma$-components)
and $f_A = f_{A+}\sigma_+ + f_{A-}\sigma_-$. Thus the complete
pseudo-complex metric is

\beqa
& \left( g_{\mu\nu} \right)  = & \nonumber \\
&
\left(
\begin{array}{cccc}
\left( 1 - \frac{2{\cal M}}{R} + \frac{\Omega_{RN}}{R} - \frac{A}{R^2} \right) e^{f_A}
& 0 & 0 & 0 \\
0 & -\left( 1 - \frac{2{\cal M}}{R} + \frac{\Omega_{RN}}{R} - \frac{A}{R^2}
\right)^{-1}
& 0 & 0 \\
0 & 0 & -R^2 & 0 \\
0 & 0 & 0 & -R^2 {\rm sin}^2\theta
\end{array}
\right)
&
\nonumber \\
\eeqa
and the projected metric is given by

\beqa
\left ( 
\begin{array}{cccc}
\left( 1 - \frac{2m}{r} + \frac{\Omega_{RN-}}{2r} -\frac{A}{r^2} \right)
e^{\frac{f_{A-}}{2}}
& 0 & 0 & 0 \\
0 & -\left( 1 - \frac{2m}{r} + \frac{\Omega_{RN-}}{2r} - \frac{A}{r^2}
\right)^{-1}
& 0 & 0 \\
0 & 0 & -r^2 & 0 \\
0 & 0 & 0 & -r^2 {\rm sin}^2\theta
\end{array}
\right)
~~~.
\nonumber \\
\eeqa

With that we obtain the length element squared, which is

\beqa
d\omega^2 & = & \left( 1 - \frac{2m}{r} + \frac{\Omega_{RN-}}{2r} - \frac{A}{r^2} \right)
e^{\frac{f_{A-}}{2}}
(dx^0)^2 \nonumber \\
& - &\left( 1 - \frac{2m}{r} + \frac{\Omega_{RN-}}{2r} - \frac{A}{r^2} \right)^{-1}
(dr)^2 - r^2 \left( (d\theta )^2 + {\rm sin}^2\theta (d\phi )^2 \right)
~~~.
\nonumber \\
\eeqa

Thus, both $g_{00}$ and $g_{11}$ do not just get a
charge dependence added as in GR, but the correction term is changed as well. Furthermore all terms of $g_{00}$ are multiplied
with a charge dependent factor $e^{\frac{f_{A-}}{2}}$. In conclusion the metric components of
the Reissner-Nordstr\"om metric are not the sum of the respective
components of the Schwarzschild metric and the simple GR charge term anymore.
Obviously, we do predict stronger deviations to GR as a priori expected.
Also note, that according to (\ref{defA}) $A$ {\it is always negative}, so that the charge prevents
the collapse to a singularity, like in standard GR.

\section{The pseudo-complex Kerr solution}

In this section we will discuss our findings concerning a pseudo-complex Kerr
solution. Some intermediate steps of the calculations can be found in the appendix
of \cite{schoenenbach}.

As previously mentioned, it is not that easy to find a pseudo-complex
Kerr solution. To do so, we will follow an ansatz first made by Carter
\cite{carter,plebanski}. He demanded the Klein-Gordon-Equation
\begin{equation}
 \frac{1}{\Psi} \fracpd{}{x^\alpha} \left( \sqrt{-g}g^{\alpha\beta} \fracpd{\Psi}{x^\beta} \right) - m_0^2\sqrt{-g} = 0
\end{equation}
to be separable. This yields an ansatz for the metric tensor
\begin{align}
 	\g &=& \frac{1}{Z} \begin{pmatrix}
	\Delta_r C_\mu^2 - \Delta_\mu C_r^2 & 0 & 0 & \Delta_\mu C_r Z_r - \Delta_r C_\mu Z_\mu \\
	0 & - \frac{Z^2}{\Delta_r} & 0 & 0 \\
	0& 0& - \frac{Z^2}{\Delta_\mu} & 0\\
	\Delta_\mu C_r Z_r - \Delta_r C_\mu Z_\mu & 0 & 0 & \Delta_r Z_\mu^2 - \Delta_\mu Z_r^2	\end{pmatrix} \quad ,
	\label{eq:ansatzcartermetrik}
\end{align}
where $, \Delta_r$ and $ Z_r$ are functions of the
variable $r$ while $\Delta_\mu$ and $ Z_\mu$ are functions of  $\mu = \cos\theta$.
$C_\mu$ and $C_r$ are constant factors, which will be determined later.
The function $Z$ is given as $Z = Z_r C_\mu - Z_\mu C_r$. Although the
metric still shows a high symmetry, it is rather tedious to compute
the Einstein equation. Some time can be saved by using a method from
differential geometry, first introduced by Cartan \cite{cartan},
to obtain the Einstein tensor. For a full explanation of this method the
reader is referred to \cite{misner,oneill,misnerpaper}.
The Einstein tensor
(remember that the Einstein equation is given by $G^\mu_\nu =
-\frac{8\pi\kappa}{c^2} T^\mu_\nu$, with
$G^\mu_\nu = {\cal R}^\mu_\nu - \frac{1}{2}g^\mu_\nu {\cal R}$)
obtained is given by
\begin{align}
 {G^0}_{0} =~&  \frac{1}{2Z} \Delta_{\mu|\mu\mu}  + \frac{1}{Z^2} \Delta_r Z_{r|rr}  + \frac{a^2}{4Z^3} \Delta_\mu \left( {Z_{\mu|\mu}}^2 +{Z_{r|r}}^2 \right)  \notag \\
& - \frac{3}{4Z^3} \Delta_r \left( {Z_{\mu|\mu}}^2 +{Z_{r|r}}^2 \right)  + \frac{a}{2Z^2}\Delta_{\mu|\mu} Z_{\mu|\mu} + \frac{1}{2Z^2} \Delta_{r|r} Z_{r|r} \notag \\
{G^0}_{3} =~ & - \frac{1}{2Z^2} \sqrt{\Delta_r \Delta_\mu} \left(a Z_{r|rr} + Z_{\mu|\mu\mu} \right) \notag \\
{G^1}_{1} =~ &  \frac{1}{2Z} \Delta_{\mu|\mu\mu} + \frac{a^2}{4Z^3} \Delta_\mu \left( {Z_{\mu|\mu}}^2 + {Z_{r|r}}^2\right) - \frac{1}{4Z^3} \Delta_r \left( {Z_{\mu|\mu}}^2 + {Z_{r|r}}^2 \right)  \notag \\
 &  + \frac{a}{2Z^2}\Delta_{\mu|\mu} Z_{\mu|\mu}  + \frac{1}{2Z^2}\Delta_{r|r} Z_{r|r} \notag \\
{G^2}_{2} =~ &   \frac{1}{2Z} \Delta_{r|rr} - \frac{a^2}{4Z^3} \Delta_\mu \left( {Z_{\mu|\mu}}^2 + {Z_{r|r}}^2 \right) + \frac{1}{4Z^3} \Delta_r \left( {Z_{\mu|\mu}}^2 + {Z_{r|r}}^2 \right) \notag \\
 &~- \frac{a}{2Z^2} \Delta_{\mu|\mu} Z_{\mu|\mu} - \frac{1}{2Z^2} \Delta_{r|r} Z_{r|r}   \notag \\
{G^3}_{3} =~ &   \frac{1}{2Z} \Delta_{r|rr} - \frac{a}{Z^2} \Delta_\mu Z_{\mu|\mu\mu}  - \frac{3a^2}{4Z^3} \Delta_\mu \left( {Z_{\mu|\mu}}^2 + {Z_{r|r}}^2 \right) \notag \\
  &~ + \frac{1}{4Z^3} \Delta_r \left( {Z_{\mu|\mu}}^2 + {Z_{r|r}}^2 \right) - \frac{a}{2Z^2} \Delta_{\mu|\mu} Z_{\mu|\mu} - \frac{1}{2Z^2} \Delta_{r|r} Z_{r|r}  \quad ,
\label{eq:einsteintensorkerr}
\end{align}
where the constant factors have been chosen as $C_r =a$,
$C_\mu=1$ \cite{carter,plebanski,editorialnotetocarter} and the subscript
$_{|\mu,r}$ stands for the derivative with respect to $\mu,r$ respectively.
As before, all calculations stay the same, when we switch to a
pseudo-complex description of the theory. Only the variational
principle changes. Thus only the $\sigma_-$ component
of the equation needs to be considered, as the $\sigma_+$ part of it is
identical to the classical Einstein equation. To solve now the Einstein
equation $G^\nu_\mu = \Xi^\nu_\mu$, we will consider similar combinations
of (\ref{eq:einsteintensorkerr}) as Pleb\'{a}nski  and Krasi\'{n}ski \cite{plebanski} did
\begin{align}
 - \frac{1}{2Z^2} \sqrt{\Delta_{R_-} \Delta_{\mu_-}} (a_- Z_{R_-|R_- R_-} + Z_{\mu_-|\mu_- \mu_-} ) = {\Xi^0}_3 \notag \\
 \frac{1}{2Z} (\Delta_{\mu_-|\mu_- \mu_-} + \Delta_{R_-|R_- R_-}) = {\Xi^1}_1 + {\Xi^2}_2 \notag \\
 \frac{a_-}{Z^2} \Delta_{\mu_-} Z_{\mu_-|\mu_- \mu_-} + \frac{a_-^2}{2Z^3} \Delta_{\mu_-} ( Z_{\mu_-|\mu_- }^2 + Z_{R_-|R_- }^2 ) = {\Xi^2}_2 - {\Xi^3}_3 \notag \\
 \frac{1}{Z^2} \Delta_{R_-} Z_{R_-|R_- R_-} - \frac{1}{2Z^3} \Delta_{R_-} ( Z_{\mu_-|\mu_- }^2 + Z_{R_-|R_- }^2 ) = {\Xi^0}_0 - {\Xi^1}_1  \notag
 \end{align}
 \begin{align}
 \frac{1}{2Z} \Delta_{R_-|R_- R_-} - \frac{a_-^2}{4Z^3} \Delta_{\mu_-} \left( {Z_{\mu_-|\mu_- }}^2 + {Z_{R_-|R_- }}^2 \right) - \frac{a_-}{2Z^2} \Delta_{\mu_-|\mu_- } Z_{\mu_-|\mu_- }  \notag \\
 + \frac{1}{4Z^3} \Delta_{R_-} \left( {Z_{\mu_-|\mu_- }}^2 + {Z_{R_-|R_- }}^2 \right) - \frac{1}{2Z^2} \Delta_{R_-|R_- } Z_{R_-|R_- }    = {\Xi^2}_2 \quad . 
~\label{eq:systemkerr}
\end{align}

We will treat the $\Xi^\nu_\mu$ as arbitrary functions in
$R_-$ and $\mu_-$ at first. This allows us to choose them properly,
so that the equations (\ref{eq:systemkerr}) can be solved.
The first step consists in setting $\Xi^3_0 = 0$ and thus the first
line in (\ref{eq:systemkerr}) becomes
\begin{equation}
 a Z_{R_-|R_- R_-} + Z_{\mu_-|\mu_- \mu_-}  = 0 \quad .
 \label{eq:ZrZmu}
\end{equation}
Choosing $\Xi_0^3 \neq 0$ would not allow an analytic solution, i.e., the
assumption  $\Xi_0^3 = 0$ is for convenience.
(\ref{eq:ZrZmu}) is formally identical to the classical case
\cite{plebanski}. We have a sum of two functions of different
variables equal to a constant. Thus both have to be constant
and we can conclude
\begin{equation}
Z_{R_-} = C R_-^2 + C_1 R_- + C_2 \quad \text{and} \quad Z_{\mu_-} = -a_- C \mu_-^2 + C_3 \mu_- + C_4 \quad.
\end{equation}

The next ad hoc choice is made with $\Xi^2_2 = \Xi^3_3$ and with the
third equation in (\ref{eq:systemkerr}) we arrive, after some manipulations,
at
\begin{equation}
 C_4 = \frac{C_2}{a_-} - \frac{C_1^2 + C_3^2}{4a_- C} \quad .
\label{eq:c4pleb}
\end{equation}
Inserted in (\ref{eq:ZrZmu}) we can observe, that the
transformation $\mu_- = \mu_-' + \frac{C_3}{2a_-C}$ together with a
redefinition $C_2 = a_- C_2' + \frac{C_1^2}{4C}$ has the same effect
as if we choose $C_3=0$ \cite{plebanski}. Thus we are left with
\begin{equation}
 Z_{R_-} = C \left( R_- + \frac{C_1}{2C} \right)^2 + a_- C_2' \quad , \quad Z_{\mu_-} = -a C\mu_-^2 + C_2' \quad .
\end{equation}

Another transformation, now for the variable $R_-$, yields the same
as if we set $C_1 = 0$. As the factor $ Z = Z_{R_-} - a_- Z_{\mu_-}$
is independent of $C_2'$ we can choose $C_2' = a_- C$ just as in
the classical case \cite{plebanski}. The final step here consist of
setting $C=1$ by redefining $\Delta_{\mu_-}$ and $\Delta_{R_-}$.
Therefore, we have determined the functions
\begin{equation}
 Z_{R_-} = R_-^2 + a_-^2 \quad , \quad Z_{\mu_-} = a_- (1-\mu_-^2)
~~~,
\label{eq:zplebanski2}
\end{equation}
which again are formally identical to the classical solution
\cite{carter,plebanski}.

Now, we consider the second equation of (\ref{eq:systemkerr}) with the
assumption $\Xi^1_1 + \Xi^2_2 = \frac{1}{2Z} \sum_{n=3}^\infty\frac{\tilde{B}_n}{R_-^n}$
(the right hand side simulates the contribution of $T^{\mu\nu}_{{\rm de}}$
of the "dark energy"),
which yields\footnote{Instead of the whole series one could choose only one or several of the terms in the series.
Later we will restrict our discussions to the case $n=3$ as in the previous sections.}
\begin{equation}
 \Delta_{{R_-}|{R_-}{R_-}} + \Delta_{{\mu_-}|{\mu_-}{\mu_-}} - \sum_{n=3}^\infty\frac{\tilde{B}_n}{R_-^n} = 0 \quad . 
\end{equation}
Again we have two functions of different variables to be equal.
This leads to
\begin{align}
 \Delta_{R_-} &= E R_-^2 - 2{\cal M}_- R_- + E_2 + \sum_{n=3}^\infty\frac{1}{(n-1)(n-2)} \frac{\tilde{B}_n}{R_-^{n-2}} \quad , \notag \\
 \Delta_{\mu_-} &= - E \mu_-^2 + E_3 \mu_- + E_4 \quad .
\end{align}
Inserting this and (\ref{eq:zplebanski2}) into the last equation of
(\ref{eq:systemkerr}) we get after some algebra
\begin{equation}
\sum_{n=3}^\infty \left(\frac{\tilde{B}_n}{R_-^{n-2}} \left( \frac{1}{n-2} + \frac{1}{2} \right) + \frac{\tilde{B}_n a^2 \mu_-^2}{2 R_-^n} \right)+ (E_2 -E_4 a^2)= Z^2 {\Xi^2}_2 \quad .
\end{equation}
If we now chose 
\begin{equation}
\Xi^2_2 = \frac{1}{Z^2}\sum_{n=3}^\infty  \left(\frac{\tilde{B}}{R_-^{n-2}} \left( \frac{1}{n-2} + \frac{1}{2} \right) + \frac{\tilde{B} a^2 \mu_-^2}{2 R_-^n} \right) \quad ,
\end{equation}
the previous equation can be fulfilled while retaining the 
condition $(E_2 -E_4 a^2) =0$ as in the classical case.

In order to determine the remaining constants in $\Delta_{R_-}$ and
$\Delta_{\mu_-}$ we will proceed analogously to  Pleb\'{a}nski  and Krasi\'{n}ski \cite{plebanski}.
At first we set $E_3 = 0$, otherwise one would have a term proportional
to $\mu_- = \cos\theta_-$, which violates the symmetry with respect to a
reflection on the equatorial plane. To avoid a coordinate singularity
at the poles, we set $E=1$. Finally we choose $E_4 = 1$ to get the
correct Schwarzschild metric in the limit $a_- \rightarrow 0$. This leaves
us then with
\begin{align}
& Z_{R_-} = R_-^2 + a_-^2 \quad , \quad Z_{\mu_-} = a_-(1-\mu_-^2) \quad , \notag \\
& \Delta_{R_-} = R_-^2 -2{\cal M}_- R_- + a_-^2 + \sum_{n=3}^\infty\frac{1}{(n-1)(n-2)} \frac{\tilde{B}_n}{R_-^{n-2}}  \quad , \notag \\
& \Delta_{\mu_-} = 1 - \mu_-^2 \quad , \quad Z = Z_{R_-} - a_- Z_{\mu_-} = R_-^2 + a_-^2 \mu_-^2 \quad .
\label{eq:loessystemkerr}
\end{align}
This can be inserted into (\ref{eq:ansatzcartermetrik}) and, together
with $\mu = \cos\theta$, we get the $\sigma_-$-part of the metric
\begin{align}
g^-_{00} &= \frac{ R_-^2 - 2{\cal M}_- R_- + a_-^2 \cos^2 \theta_- + \sum_{n=3}^\infty\frac{1}{(n-1)(n-2)} \frac{\tilde{B}_n}{R_-^{n-2}} }{R_-^2 + a_-^2 \cos^2\theta_-} \notag \\
g^-_{11} &= - \frac{R_-^2 + a_-^2 \cos^2 \theta_-}{R_-^2 - 2{\cal M}_- R_- + a_-^2 + \sum_{n=3}^\infty\frac{1}{(n-1)(n-2)} \frac{\tilde{B}_n}{R_-^{n-2}}} \notag \\
g^-_{22} &= - R_-^2 - a_-^2 \cos^2 \theta_-  \notag \\
g^-_{33} &= - (R_-^2 +a_-^2 )\sin^2 \theta_- - \frac{a_-^2 \sin^4\theta_-\left(2{\cal M}_- R_- - \sum_{n=3}^\infty\frac{1}{(n-1)(n-2)} \frac{\tilde{B}_n}{R_-^{n-2}} \right)}{R_-^2 + a_-^2 \cos^2 \theta_-}  \notag \\
g^-_{03} &= \frac{-a_- \sin^2 \theta_- ~2{\cal M}_- R_- + a_- \sum_{n=3}^\infty\frac{1}{(n-1)(n-2)} \frac{\tilde{B}_n}{R_-^{n-2}}  \sin^2 \theta_- }{R_-^2 + a_-^2 \cos^2\theta_-}   \quad .
\label{sol-kerr}
\end{align}
Note, that in spite of all assumptions made, (\ref{sol-kerr})
represents a new Kerr solution also in standard GR with a special
$T^{\mu\nu}$ tensor. This is because, as mentioned
before, each $\sigma_\pm$ component describes one particular GR with a given
symmetric metric.

The $\sigma_+$-component matches the classical Kerr solution.
Finally, projecting the pc-metric on its real part, as described above,
yields the metric
\begin{align}
 g^{\text{Re}}_{00} &= \frac{ r^2 - 2m r  + a^2 \cos^2 \theta + \sum_{n=3}^\infty\frac{1}{(n-1)(n-2)} \frac{\tilde{B}_n}{2r^{n-2}} }{r^2 + a^2 \cos^2\theta} \notag \\
g^{\text{Re}}_{11} &= - \frac{r^2 + a^2 \cos^2 \theta}{r^2 - 2m r + a^2 + \sum_{n=3}^\infty\frac{1}{(n-1)(n-2)} \frac{\tilde{B}_n}{2r^{n-2}} } \notag \\
g^{\text{Re}}_{22} &= - r^2 - a^2 \cos^2 \theta  \notag \\
g^{\text{Re}}_{33} &= - (r^2 +a^2 )\sin^2 \theta - \frac{a^2 \sin^4\theta \left(2m r - \sum_{n=3}^\infty\frac{1}{(n-1)(n-2)} \frac{\tilde{B}_n}{2r^{n-2}} \right)}{r^2 + a^2 \cos^2 \theta}  \notag \\
g^{\text{Re}}_{03} &= \frac{-a \sin^2 \theta ~ 2m r + a \sum_{n=3}^\infty\frac{1}{(n-1)(n-2)} \frac{\tilde{B}_n}{2r^{n-2}}  \sin^2 \theta }{r^2 + a^2 \cos^2\theta}   \quad . 
\label{eq:kerrpseudo1}
\end{align}
\bigskip \\

For the further discussions we will only consider the case $n=3$, which 
results in the metric
\begin{align}
 g^{\text{Re}}_{00} &= \frac{ r^2 - 2m r  + a^2 \cos^2 \theta + \frac{B}{2r} }{r^2 + a^2 \cos^2\theta} \notag \\
g^{\text{Re}}_{11} &= - \frac{r^2 + a^2 \cos^2 \theta}{r^2 - 2m r + a^2 +  \frac{B}{2r}  } \notag \\
g^{\text{Re}}_{22} &= - r^2 - a^2 \cos^2 \theta  \notag \\
g^{\text{Re}}_{33} &= - (r^2 +a^2 )\sin^2 \theta - \frac{a^2 \sin^4\theta \left(2m r -  \frac{B}{2r}  \right)}{r^2 + a^2 \cos^2 \theta}  \notag \\
g^{\text{Re}}_{03} &= \frac{-a \sin^2 \theta ~ 2m r + a \frac{B}{2r}   \sin^2 \theta }{r^2 + a^2 \cos^2\theta}   \quad . 
\label{eq:kerrpseudo}
\end{align}

As this metric (\ref{eq:kerrpseudo}) represents the pseudo-complex 
equivalent to the Kerr solution, it is of interest, wether one can still 
identify the parameter $a$ with the angular momentum $J$. To see that
this identification still holds, we will follow Adler 
et al. \cite{adler} and expand the line element given by (\ref{eq:kerrpseudo})
linear in $a$
\begin{align}
 \dd s^2 = &\left( 1- \frac{2m}{r} + \frac{B}{2 r^3} \right) \dd t^2 - \frac{1}{1- \frac{2m}{r} + \frac{B}{2 r^3}} \dd r^2 - r^2 \dd \theta^2 - r^2 \sin^2\theta \dd \phi^2 \notag \\
& +2 a \sin^2 \theta \left(- \frac{2m}{r} +\frac{B}{2r^3} \right) \dd \phi \dd t  \quad .
\label{eq:kerrentwickelt1}
\end{align} 

This expansion represents the limit of a slowly rotating body.
Next we expand (\ref{eq:kerrentwickelt1}) linear in $\frac{1}{r}$. This 
is the limit for large distances. The line element now takes the form
\begin{align}
  \dd s^2 = &\left( 1- \frac{2m}{r}  \right) \dd t^2 - \left(1+ \frac{2m}{r} \right) \dd r^2 - r^2 \dd \theta^2 - r^2 \sin^2\theta \dd \phi^2 \notag \\
& -2 a \sin^2 \theta \frac{2m}{r}   \dd \phi \dd t \quad .
\label{eq:kerrentwickelt2}
\end{align}
In the following discussion we will focus on the term 
proportional to $\dd \phi \dd t$.
\begin{equation}
 -2 a \sin^2\theta \frac{2m}{r} \dd \phi \dd t \quad .
\end{equation}
To compare this term 
with the calculations of Lense and Thirring
\cite{adler,lensethirring} 
\begin{align}
 \dd s^2 =& \left(1 - \frac{2m}{\rho} \right) \dd t^2 - \left( 1+ \frac{2m}{\rho} \right)\left[ \dd \rho^2 + \rho^2 \left( \dd \theta^2 + \sin^2 \theta \dd \phi^2 \right)\right] \notag \\
& + \frac{4 \kappa J}{c^3 \rho} \sin^2\theta \dd \phi \dd t \quad ,
\end{align}
we note, that for large distances the
two radial variables $r$ and $\rho$ (isotropic coordinates, 
used by Lense and Thirring) coincide. This finally yields
the same connection between $a$ and the angular
momentum $J$ as in the classical case
\begin{equation}
  a = -  \frac{\kappa J}{m c^3} \quad.
\label{eq:andJ}
\end{equation}

Obviously, the parameter $a$ can still be identified with
the angular momentum of the source.

\vskip 0.5cm

The classical Kerr solution shows some special hypersurfaces which are
of great physical interest. One of these we already know from the
classical Schwarzschild solution: In the orbital plane the radius of
this sphere is at
$r = 2m$, a sphere with infinite red shift. In \cite{hess1}
the specific choice $B > 2 m^2$ had the effect, that this infinite
red shift surface and the singularity at the origin vanished. For
corrections proportional to $\frac{B}{2r^3}$ this happens for
$B > (4/3m)^3$ (see (\ref{eq:bedingungfuerB})). We will now investigate the influence of the
additional term proportional to $B$ on the existence of an event
horizon for the Kerr metric.

As shown in \cite{adler} surfaces corresponding to $g_{00} = 0$
can be passed in both directions by an observer (except at the poles),
e.g. these surfaces are no event horizons. 

The property of a surface to be an event horizon can be investigated
through the norm of its normal vector $n_\alpha$. Only if it is negative,
physical observers can pass in both directions. A normal vector with positive
norm corresponds to a timelike surface. Such surfaces can only be
passed in one direction by physical observers. \\
Now we will look for time independent axially symmetric surfaces with a
null normal vector. These surfaces can be described by \cite{adler}
\begin{equation}
 u(r,\theta) = \text{const} \quad .
 \label{eq:surfaces}
\end{equation}
Their normal vector is then given by
\begin{equation}
  n_\alpha = \left(0 , \fracpd{u}{r} , \fracpd{u}{\theta}, 0 \right) \quad.
\end{equation}
Setting the norm $n_\alpha n^\alpha = 0$ yields the equation
\begin{equation}
 \left(r^2 - 2m r + a^2 + \frac{B}{2r} \right)  \left( \fracpd{u}{r}\right)^2 + \left( \fracpd{u}{\theta}\right)^2 = 0 \quad ,
\end{equation}
which can be solved by a product ansatz $u = R(r) \Theta(\theta)$
\begin{equation}
- \left(r^2 - 2m r + a^2 + \frac{B}{2r} \right)  \left( \frac{\fracpd{R}{r}}{R}\right)^2 = \left( \frac{\fracpd{\Theta}{\theta}}{\Theta}\right)^2 \quad .
\end{equation}
Both sides of this equation depend on different variables and thus have to
be constant. In analogy to \cite{adler} we will call that constant
$\lambda$ which then gives
\begin{equation}
 \Theta = A e^{\sqrt{\lambda} \theta} \quad.
\end{equation}

This expression however is not periodic in $\theta$ and
therefore can't describe a surface except for the case where
$\lambda = 0$, which means $\Theta = \text{const}$. The remaining
equation for $R$ then is
\begin{equation}
 \left(r^2 - 2m r + a^2 + \frac{B}{2r} \right)  \left( \frac{\fracpd{R}{r}}{R}\right)^2 = 0 \quad.
\end{equation} 
Excluding the trivial case $\fracpd{R}{r} = 0$ we are
left with the solution of
\begin{equation}
 \left(r^2 - 2m r + a^2 + \frac{B}{2r} \right) = 0 \quad .
\label{eq:rootskerr}
\end{equation}
Possible physical solutions for $r$ are given by positive real roots of the cubic polynomial 
\begin{equation}
p(r)= r^{3}-2mr^{2}+a^{2}r+\frac{B}{2}\quad.
\end{equation}
Since the derivative
\begin{equation}
p'(r)=3r^{2}-4mr+a^{2}\quad
\end{equation}
is positive for all $r\leq 0$, from $p(0)=B/2>0$ and $\lim_{r\to-\infty}p(r)=-\infty$ it follows that $p(r)$ has always exactly one negative real root, which is not relevant for our argument. Depending on the parameters $a^{2}$ and $B$ there might be two more real roots, which then have to be positive numbers and thus would represent possible solutions of (\ref{eq:rootskerr}). It is well known that a cubic function has three distinct real roots if it has a positive discriminant\cite{abramowitz}. For $p(r)$ the parameter dependent discriminant $D(a^2,B)$ reads
\begin{equation}
D(a^2,B)=\frac{1}{27}\left(4\left(4m^2-3a^2\right)^3 - \left(18ma^2-16m^3+\frac{27}{2}B\right)^2\right)\quad.
\end{equation}
It is easy to see that a first condition for $D(a^2,B)>0$ is already given by $a^2<(4/3)m^2$. We rewrite the condition $D(a^2,B)>0$ by use of the parametrization $a^2=\epsilon(4/3) m^2$, with $\epsilon\in[0,1]$, and obtain
\begin{equation}
\label{eq:discriminant2}
4\left(4(1-\epsilon)m^2\right)^3 > \left(8m^3(3\epsilon-2)+\frac{27}{2}B\right)^2\quad.
\end{equation}
Now we determine the maximum parameter value $B^{*}$ for which this condition can be satisfied. The left hand term monotonically decreases with increasing $\epsilon$, whereas the right hand term monotonically increases as long as the term in the bracket is positive. If for some $\epsilon$ and $B$ the condition is met with a negative term in the bracket on the right hand side, we can choose a larger $B$ such that this term is positive and the condition is still fulfilled. It follows that the maximum value $B^{*}$ satisfying the condition (\ref{eq:discriminant2}) is obtained for $\epsilon=0$, or equivalently $a=0$. In this case (\ref{eq:discriminant2}) reads
\begin{equation}
4\left(4m^{2}\right)^{3} = \left(16m^3\right)^2> \left(\frac{27}{2}B-16m^{3}\right)^2\quad ,
\end{equation}
which yields $B^{*}=(4/3)^3 m^3$. This value corresponds to the limiting case for the Schwarzschild solution including corrections proportional to $\frac{B}{2r^3}$ as discussed at the end of chapter 3 (see (\ref{eq:bedingungfuerB})). We conclude that also for $a>0$, for $B>B^*$ there are no positive real roots of (\ref{eq:rootskerr}) and therefore just as in the Schwarzschild case the modified Kerr
solution shows no event horizons. Note that there also are no
surfaces of infinite redshift as $g_{00}= 0$ is included in our discussion
of (\ref{eq:rootskerr}) because $a^2$ and $a^2 \cos^2\theta$ have the 
same range of values.

\section{Some experimental considerations}

In this section we discuss the phenomenon, that a light emitting blob of plasma
is circulating an {\it Active Galactic Nucleus} (AGN). These objects are
observed and discussed in the literature \cite{boller,genzel,eckart}. Orbiting around
the AGN, near to the Schwarzschild radius, they appear at a well determined
frequency. Using standard GR for the Kerr solution one obtains from the orbiting frequency
the distance $r$ of this plasma blob from the center.
This assumes standard GR. The deduced $r$ value will change, of course, when
pc-GR is applied. Here, we
will determine the frequency as a function of the radial distance for the
case of a black hole in Einstein's GR, for the pc-Schwarzschild solution and
the same for the Kerr solution. Differences will be pointed out, in particular such
signals, which, eventually, can not be explained by standard GR, may serve as a possible
sign for pc-GR.

In a first step, we consider a non-rotating gray star. The real situation
would be a rotating gray star, which we will discuss
further below. The Schwarzschild case serves for illustrating purposes.

Furthermore, we restrict
to a simplified circular orbit. The Lagrange function in standard GR is given by\footnote{We neglect in the further discussions the factor $e^{\frac{f_-}{2}}$, assuming that it is small.}

\begin{equation}
 L =  g_{00}~ c^2\dot{t}^2 + g_{11} \dot{r}^2
 - r^2  [ \dot{\theta}^2 + \sin^2\theta \dot{\phi}^2 ] =
 \frac{ds^2}{ds^2} = 1
 \label{Lagrange}
\end{equation}
The dot indicates the derivative with respect to $s$.
Using the variational principle for the geodesic equations, we arrive at the
following equations of motion

\begin{align}
\frac{d}{ds} ( g_{00}c^2 \dot{t} ) & = 0 \nonumber \\
  \frac{d}{ds} \left( 2 g_{11} \dot{r} \right) & =
  \fracpd{g_{00}}{r} c^2\dot{t}^2 +
  \fracpd{g_{11}}{r} \dot{r}^2 -2 r[ \dot{\theta}^2 +
  \sin^2\theta \dot{\phi}^2 ] \nonumber \\
\frac{d}{ds} ( -2 r^2 \dot{\theta}) + 2 r^2 \sin\theta\cos\theta
\dot{\phi}^2 & = 0 \nonumber \\
\frac{d}{ds} (-2 r^2 \sin^2\theta \dot{\phi}) & = 0
\label{geod}
\end{align}
The structure will not be much different, when we move to the pseudo-complex description.
The change consists in modified expressions for the metric components.

In the Schwarzschild case, we can restrict to a circular motion with
$\theta = \frac{\pi}{2}$. Then the former equations simplify to

\begin{align}
\frac{d}{ds} ( g_{00} c^2\dot{t} ) & = 0 \nonumber \\
  \frac{d}{ds} \left( g_{11}\dot{r} \right) &=
  \fracpd{g_{00}}{r} c^2\dot{t}^2 +
  \fracpd{g_{11}}{r} \dot{r}^2 -2 r \dot{\phi}^2  \nonumber \\
\frac{d}{ds} (-2 r^2 \dot{\phi}) & = 0
\label{geod-simplyfied}
\end{align}
From the third equation follows the conservation of angular momentum

\beqa
 H = r^2 \dot{\phi} = r^2 \frac{d\phi}{dt} \dot{t} = r^2 \omega \dot{t} = \text{const}
~~~,
\eeqa
where $H$ is proportional to the angular momentum.

From the first equation in (\ref{geod-simplyfied}) the derivative of the time
with respect to the eigentime (the derivative is indicated by a dot)
is determined, i.e.,

\beqa
 2g_{00}c^2 \dot{t} &= 2Ac \nonumber \\
\dot{t} &= \frac{A}{g_{00}c} \label{eq:tdotA}
\eeqa

Restricting to circular orbits, we have $r=r_0$ and ${\dot r}=0$. Due to this, the
integration constant can be related to the frequency $\omega$ of the circular
motion. To do that we insert $\theta = \frac{\pi}{2}$, $\dot{\phi} = \omega \dot{t}$ and $\dot{r} = \dot{\theta}= 0$ 
into equation (\ref{Lagrange}) which then becomes
\begin{equation}
  g_{00}~ c^2\dot{t}^2  - r^2   \dot{\phi}^2  =  1 \quad \Rightarrow \quad g_{00} ~ c^2 \dot{t}^2 -r^2 \omega^2 \dot{t}^2 = 1 \quad  \Leftrightarrow \quad \dot{t}^2= \frac{1}{g_{00} c^2 - r^2 \omega^2} \quad .
\end{equation}
Now we can use (\ref{eq:tdotA}) and obtain
\beqa
 A = \frac{g_{00}c}{\sqrt{g_{00}c^2-r^2\omega^2}}
~~~.
\eeqa
The second equation in (\ref{geod-simplyfied}) simplifies to

\begin{equation}
  0 = \fracpd{g_{00}}{r}c^2 \dot{t}^2  - 2 r \dot{\phi}^2
  ~~~.
\end{equation}
Considering that ${\dot \phi}
= \frac{d\phi}{ds} = \frac{d\phi}{dt} \frac{dt}{ds} = \omega \dot{t}$
we obtain

\begin{equation}
 0 = \fracpd{g_{00}c^2}{r} -2 r \omega^2 ~~ \Leftrightarrow ~~
 \omega = \pm c \sqrt{\frac{1}{2r}\fracpd{g_{00}}{r} }
~~~.
\label{final-omega-schw}
\end{equation}

This expression is valid for both the GR and for the pc-GR Schwarzschild case. Using the
$g_{00}$ component for standard GR and then for pc-GR (assuming a correction
to the metric of $\frac{B}{2r^3}$, see the end of chapter 3.2), we
obtain respectively

\beqa
\omega_{{\rm GR}} & = & c \sqrt{\frac{m}{r^3}} \nonumber \\
\omega_{{\rm pc-GR}} & = & \sqrt{ \frac{m}{r^3} - \frac{3B}{4r^5}}
~~~.
\label{frec-schw}
\eeqa

These are two different relations, which yield for a fixed, observed frequency
two different radial distances.

\begin{figure}[ht]	
\begin{center}
\begin{center}
\rotatebox{270}{\resizebox{250pt}{350pt}{\includegraphics[width=0.5\textwidth]{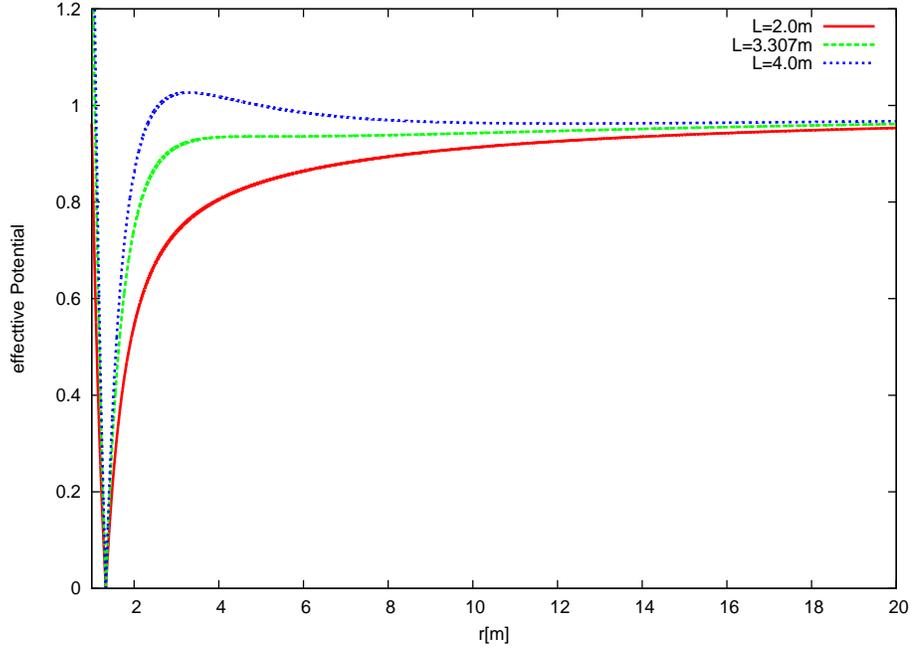}}}
\end{center}
\begin{flushleft}
\caption{
The effective potential for the pc-Schwarzschild solution. The value $L=3.307m$ corresponds
to the last stable orbit (saddle point at $r\approx 5.25$). \label{fig:EffPotSchwarz}
}
\end{flushleft}
\end{center}
\end{figure}

In the pc-GR there is also a {\it last stable orbit}. The reason for that is
the structure of the effective potential, which analogous to Misner et al. (page 639) \cite{misner}
has the form

\beqa
V^2 & = & \left(1-\frac{2m}{r}+\frac{B}{2r^3}\right) \left[ 1
+ \frac{L^2}{r^2} \right]
~~~.
\label{V-eff-1}
\eeqa
Choosing as $B=\frac{64}{27}m^3$, the first factor in (\ref{V-eff-1})
becomes zero at two thirds of the Schwarzschild radius (see Fig. \ref{fig:EffPotSchwarz}). For this limit, the
behavior for larger $r$ is the same as in the standard theory. \\

A local minimum exists for $r>\frac{4m}{3}$ and vanishes at approximately
$r=5.25~m$. This means that the last stable orbit is at a slightly smaller
position than in standard GR, where it is at $r=6m$, three times the
Schwarzschild radius. Thus, one possibility is to look for orbits which
exist at smaller distances than $6m$, if the gray star does not rotate.

There is another consequence: The maximal observable velocity of these
blobs is given by $\omega r$. In case of GR and pc-GR respectively the 
maximal velocity is given by

\beqa
v^{GR}_{{\rm max}} & = & c \sqrt{\frac{m}{6m}} \lesssim 0.409 c \nonumber \\
 v^{pc-GR}_{{\rm max}} & = &
 \pm c \sqrt{\frac{m}{r} - \frac{3B}{4r^3}}
 < c \sqrt{\frac{m}{r} - \frac{16m^3}{9r^3}} \nonumber \\
 & < & 0.423c
 ~~~,
\eeqa
where in the last relation we used for $r$ the value at the last stable
orbit at $5.25 m$. Obviously in both cases the maximal velocities are nearly equal.

Of course, this is only valid
for a non-rotating large central mass. In a real situation one has to study
the Kerr solution, which we will consider next:

The calculations follow along the same lines as for the non-rotating large mass.
The Lagrange function is given by

\beqa
L & = & g_{00}c^2 \dot{t}^2 + g_{11} c^2 \dot{r}^2 +
g_{22} \dot{\theta}^2 + g_{33} \dot{\phi}^2
+ 2g_{03} c\dot{t}\dot{\phi}
~~~.
\eeqa

As in the Schwarzschild case, the equation of motion for the radial part
is used to determine the orbital frequency, i.e.,

\beqa
 \frac{d}{ds} \left ( 2g_{11}\dot{r} \right ) &= \fracpd{g_{00}}{r}
 c^2\dot{t}^2 + \fracpd{g_{11}}{r} \dot{r}^2 + \fracpd{g_{22}}{r}
 \dot{\theta}^2 + \fracpd{g_{33}}{r}
 \dot{\phi}^2 + 2\fracpd{g_{03}}{r} c\dot{t}\dot{\phi} \nonumber \\
&=: g^\prime_{00} c^2\dot{t}^2 + g^\prime_{11} \dot{r}^2 +
g^\prime_{22} \dot{\theta}^2 + g^\prime_{33} \dot{\phi}^2 +
2g^\prime_{03} c\dot{t}\dot{\phi}
~~~,
\eeqa
where the prime indicates a partial derivative with respect to the radial
distance.
We restrict to orbital motion in the horizontal plane, with
$r=r_0$, ${\dot r} = 0$, $\theta = \frac{\pi}{2}$ and ${\dot \theta}=0$.
With this, the last equation simplifies to

\beqa
0&= g^\prime_{00}(r_0) c^2\dot{t}^2 + g^\prime_{33}(r_0) \dot{\phi}^2
+ 2g^\prime_{03}(r_0) c\dot{t}\dot{\phi} \nonumber \\
&= g^\prime_{00}(r_0) c^2\dot{t}^2 + g^\prime_{33}(r_0)
\omega^2\dot{t}^2 + 2g^\prime_{03}(r_0) \omega c\dot{t}^2
~~~.
\eeqa
and the orbital frequency results as

\beqa
\omega_{\pm}^{{\rm pc-GR}} & = & -c\frac{g^\prime_{03}}{g^\prime_{33}} \pm
c\sqrt{\left (\frac{g^\prime_{03}}{g^\prime_{33}}\right )^2 -
\frac{g^\prime_{00}}{g^\prime_{33}}}
~~~.
\eeqa

Substituting the metric components for the pc-Kerr-problem (see eqs. (\ref{eq:kerrpseudo}),(\ref{eq:andJ}))

\begin{align}
g_{00} &= 1 - \frac{2m}{r} + \frac{B}{2r^3} \nonumber \\
g_{03} &= \frac{-2a m r + a\frac{B}{2r}}{r^2} =
a \left ( -\frac{2m}{r} + \frac{B}{2r^3} \right ) \nonumber \\
g_{33} &= - \left ( r^2 + a^2 \right ) - a^2 \left (
\frac{2m}{r} - \frac{B}{2r^3} \right )
~~~,
\end{align}

we obtain finally for the orbital frequency

\beqa
\omega_{\pm}^{{\rm pc-GR}} & = & \frac{ac \left ( \frac{2m}{r^2} - \frac{3B}{2r^4}\right )}{2r - a^2 \left ( \frac{2m}{r^2} - \frac{3B}{2r^4} \right)} \pm \frac{c\sqrt{\frac{4m}{r}-\frac{3B}{r^3}}}{\left | 2r - a^2 \left ( \frac{2m}{r^2} - \frac{3B}{2r^4} \right ) \right |}
~~~. \label{eq:kreisfrequenz}
\eeqa
The standard GR solution is obtained by setting $B=0$. If, as usual, $a>0$ is
assumed, then the negative sign corresponds to an orbital motion in phase
with the rotation of the gray star and the positive sign is for
a motion opposite to its rotation \footnote{The parameter $a$ sometimes is defined with an opposite sign like in the book of Misner et al. \cite{misner}.}.

\begin{figure}[ht]	
\begin{center}
\begin{center}
\rotatebox{270}{\resizebox{250pt}{350pt}{\includegraphics[width=0.5\textwidth]{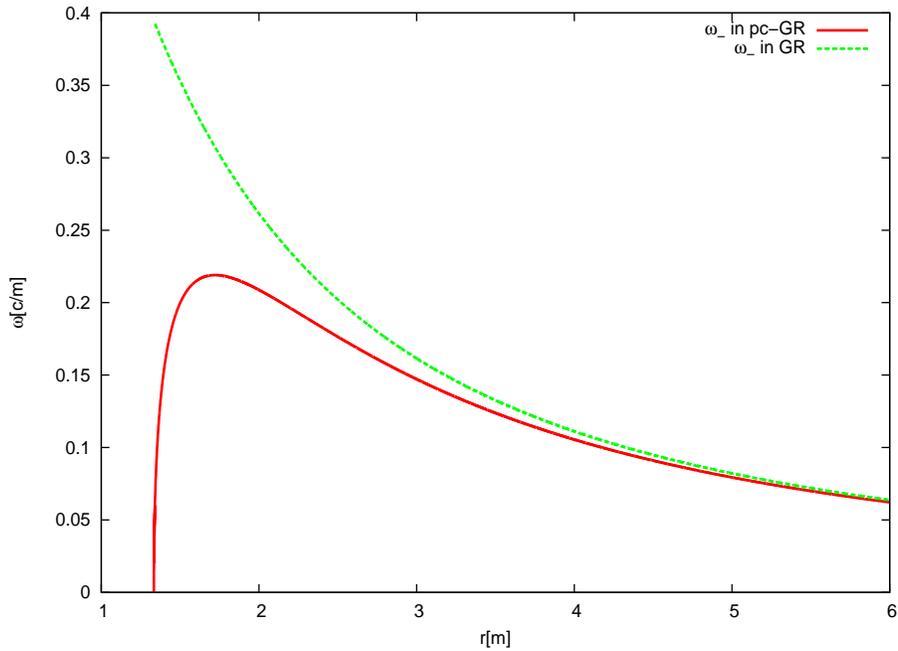}}}
\end{center}
\begin{flushleft}
\caption{The orbital frequency of a co-rotating object in a stable orbit versus the radial distance $r$ for $a=0.995$.
The plot for $\omega$ in GR starts at the last stable orbit, which is $r= 1.341m$. For $\omega$ in pc-GR
it starts at $r = \frac{4}{3}m$ - for radii below this value equation (\ref{eq:kreisfrequenz}) has no real 
solutions anymore. Thus we do not expect circular geodesic orbits below this value of $r$.
The $m$ is given by $\frac{\kappa M}{c^2}$, with $\kappa$ being the gravitational
constant. Sometimes $m$ is denoted as $r_g$ in literature \protect\cite{genzel,eckart}. \label{fig2}}
\end{flushleft}
\end{center}
\end{figure}

In Fig. \ref{fig2} the orbital frequency $\omega =2\pi \nu$ is
plotted versus $r$, for an $a=0.995$. The units are given in $\frac{c}{m}$. Using
the literature values for the gravitational constant ($6.674 \cdot 10^{-11} m^3~ kg^{-1}~ s^{-2}$), the speed of light
($3 \cdot 10^8 m ~s^{-1}$) and the mass of the sun ($2 \cdot 10^{30} kg$), we obtain

\beqa
\frac{c}{m} & \approx & 1.22 \times 10^{7} \frac{M_{{\rm sun}}}{M}
~{\rm min}^{-1}
~~~.
\eeqa
The unit is in one over minutes, $M_{{\rm sun}}$ is the mass of the
sun and $M$ is the mass of the central object.

Taking into account that the mass of the object in Sagittarius A,
the center of our galaxy, is of the order of
$3 \times 10^7M_{{\rm sun}}$, we get

\beqa
\frac{c}{m} & \approx & 3.3~{\rm min}^{-1}
~~~.
\eeqa

$\omega$ is given by $2\pi \nu$, i.e., for $\nu$ we obtain, using the maximum value
of $\omega \approx 0.219$,
$\nu = 0.115~{\rm min}^{-1}$. This corresponds to the minimal time of $8.7~{\rm min}$ for
a mass to circulate around the large mass.
Fig. \ref{fig2} indicates that the standard GR
produces larger frequencies, thus a lesser time to circulate
the large mass, which should be measurable. The pc-Kerr solution exhibits a maximum frequency,
which is due to the change of sign in the second derivative of
$g_{00}$. This property implies a minimal time of orbit, which would not exist in standard GR.
This should be a clear sign to distinguish between GR and pc-GR.

\section{Conclusions}

In this paper we revisited the pseudo-complex General Relativity,
as proposed by P.O. Hess and W. Greiner \cite{hess1}. Inconsistencies in the projection to real results
were found and discussed. A corrected projection rule was
presented.

The new procedure was applied to the case of a non-rotating gray star, the
pseudo-complex Schwarzschild solution.

In the last two chapters new pseudo-complex solutions were constructed,
namely the pseudo-complex Reissner-Nordstr\"om solution for a charged
gray star and the pseudo-complex Kerr solution for a rotating
gray star. The calculational procedures were rather complex,
nevertheless analytic solutions were found.

In all cases, the modified variational principle introduced contributions,
which can be interpreted as dark energy, acting repulsively such that
the formation of an event horizon and a singularity at the center is
avoided. This is a most important result, as any proper theory should
not contain singularities.

The origin of the dark energy stems from different field equations (different
with respect to Einstein's GR). This again, is most satisfying: Dark energy can be
introduced by modified field equations in the pseudo-complex treatment of general relativity.

In section 2 we also commented on the corrections to the theory,
when the contributions of the minimal length scale are included and their
consequences, resulting in the dispersion relation of a particle.
As explained, because of the change in the four-velocity is 
considered to be small (small acceleration), the contribution
of the minimal length can savely
be neglected. Nevertheless, in future one should investigate
the contributions of the minimal length when the acceleration
approaches $1/l$. An eight-dimensional formulation ($X^\mu = x^\mu + I \frac{l}{c} u^\mu$)
should be systematically worked out!

Finally, in the last section, we discussed observable effects leading
to differences between standard (Einstein's) and pseudo-complex General
Relativity. 
We concentrated on circular orbits around
a gray star. The pc-Schwarzschild case, of a non-rotationg
gray star, and the pc-Kerr case, of a rotating gray star,
were investigated. In case of the pc-Schwarzschild solution the last stable
orbit changed from $6m$ to  about $5.25m$. In the pc-Kerr solution
the last stable orbit is further out and the orbital frequency
is always lower than in the standard Schwarzschild case, exhibiting
also a maximum value, i.e., a minimal time for the orbit.

\section*{Acknowledgments}
The authors express sincere gratitude for the possibility to work at
the {\it Frankfurt Institute for Advanced Studies} with the excellent
working atmosphere encountered there. They acknowledge fruitful discussions
with Prof. Thomas Boller from the "Max-Planck-Institut f\"ur extraterrestrische Physik, 
Garching" and Dr. Andreas M\"uller from the "Exzellenzcluster 'Origin and 
Structure of the Universe', Garching".
Peter Otto Hess also acknowledges financial support from {\it Direcci\'{o}n General Asuntos del Personal Acad\'{e}mico} (DGAPA)
and {\it Consejo Nacional de Ciencia y Tecnolog\'{i}a } (CONACyT). Thomas Sch\"onenbach acknowledges financial support
from {\it Stiftung Polytechnische Gesellschaft Frankfurt am Main}. Gunther Caspar acknowledges financial support from {\it Frankfurt Institute for Advanced Studies}.
                                                                                            
\newpage
\begin{center}
\huge Appendix
\end{center}
\section*{Reissner-Nordstr\"om solution with unchanged source}

Assuming that $\Xi_\mu^{RN} = \Xi_\mu$, where the superscript "RN" refers to Reissner-Nordstr\"om, equation (\ref{eq-3}) leads to
\begin{align}
e^{-\lambda_{RN}} \xi^{RN}_0 &= e^{-\lambda} \xi_0 \nonumber\\
e^{-\lambda_{RN}} \xi^{RN}_1 &= e^{-\lambda} \xi_1 \nonumber\\
\xi^{RN}_2 &= \xi_2
~~~, \label{conxi}
\end{align}
so that
\beqa
\xi^{RN}_0 - \xi^{RN}_1 = e^{\lambda_{RN-} - \lambda_-} \left (\xi_0 - \xi_1 \right )
\eeqa
follows.\\
Combining this with (\ref{RNg11}) results in
\beqa
e^{-\lambda_{RN-}} = e^{-\lambda_-} - \frac{A}{R_-^2} \label{eq:lambdaRN}
\eeqa
In the next step, we have to use (\ref{diffeqRN}), but since the equation was not explicitly proven within the main chapter we will do that first. Therefore we take (\ref{RNdeq}) and replace $\nu^\prime_{RN-}$ and $\nu^{\prime\prime}_{RN-}$ by using the equations (\ref{RNxi0-xi1}) and (\ref{RNnu-pp}), which gives
\beqa
& -\lambda^{\prime\prime}_{RN-} + \frac{1}{2} \left ( \xi^{RN}_0 - \xi^{RN}_1 \right ) + \frac{R_-}{2} \left ( \xi^{RN\prime}_0 - \xi^{RN\prime}_1 \right ) \nonumber \\
& - \frac{\lambda^{\prime}_{RN-}}{2} \left ( - \lambda^{\prime}_{RN-} + \frac{R_-}{2} \left ( \xi^{RN}_0 - \xi^{RN}_1 \right ) \right ) & \nonumber\\
&+ \frac{1}{2} \left ( - \lambda^{\prime}_{RN-} + \frac{R_-}{2} \left ( \xi^{RN}_0 - \xi^{RN}_1 \right ) \right )^2 - \frac{2\lambda_{RN-}^{\prime}}{R_-} &
\nonumber \\
& = &
\nonumber \\
& \xi^{RN}_1 - \frac{2A}{R_-^4} e^{\lambda_{RN-}} &
~~~.
\nonumber \\
\eeqa  
In the next step we eliminate the big brackets
\beqa
& -\lambda^{\prime\prime}_{RN-} + \frac{1}{2} \left ( \xi^{RN}_0 - \xi^{RN}_1 \right ) + \frac{R_-}{2} \left ( \xi^{RN\prime}_0 - \xi^{RN\prime}_1 \right ) + \left (\lambda^{\prime}_{RN-} \right )^2 & \nonumber\\
&- \frac{3}{4} R_- \lambda^{\prime}_{RN-} \left ( \xi^{RN}_0 - \xi^{RN}_1 \right ) + \frac{1}{8} R_-^2 \left ( \xi^{RN}_0 -\xi^{RN}_1 \right )^2 - \frac{2\lambda_{RN-}^{\prime}}{R_-} &
\nonumber \\
& = &
\nonumber \\
& \xi^{RN}_1 - \frac{2A}{R_-^4} e^{\lambda_{RN-}} &
\nonumber \\
\eeqa
and rearrange the terms
\beqa
& -\lambda^{\prime\prime}_{RN-} + \frac{1}{2} \left ( \xi^{RN}_0 - \xi^{RN}_1 \right ) + \frac{R_-}{4} \left ( \xi^{RN\prime}_0 - \xi^{RN\prime}_1 \right ) + \left (\lambda^{\prime}_{RN-} \right )^2 & \nonumber\\
&- \frac{1}{4} R_- \lambda^{\prime}_{RN-} \left ( \xi^{RN}_0 - \xi^{RN}_1 \right ) - \frac{2\lambda_{RN-}^{\prime}}{R_-} + \frac{2A}{R_-^4} e^{\lambda_{RN-}} - \xi^{RN}_1 &
\nonumber \\
& = &
\nonumber \\
& -\frac{1}{4}R_-\left( \xi^{RN\prime}_0-\xi^{RN\prime}_1 \right)
+\frac{1}{2}\lambda^{RN\prime}_- R_- \left( \xi^{RN}_0-\xi^{RN}_1\right)
-\frac{1}{8}R_-^2 \left( \xi^{RN}_0-\xi^{RN}_1\right)^2 &
\nonumber \\
~~~.
\eeqa
When we differentiate (\ref{RNg11dif}) and reorganize the terms in a way, so that $\xi_2^{\prime}$ stands alone on the left side, we get
\begin{align}
\xi_2^{\prime} &= \left ( R_- e^{-\lambda_{RN-}} \right )^{\prime\prime} + \frac{1}{2} R_- e^{-\lambda_{RN-}} \left ( \xi^{RN}_0 - \xi^{RN}_1 \right ) \nonumber\\
&~- \frac{1}{4} R_-^2 \lambda_{RN-}^\prime e^{-\lambda_{RN-}} \left ( \xi^{RN}_0 - \xi^{RN}_1 \right ) + \frac{1}{4} R_-^2 e^{-\lambda_{RN-}} \left ( \xi^{RN\prime}_0 - \xi^{RN\prime}_1 \right ) + \frac{2A}{R_-^3}\nonumber\\
&= -2\lambda_{RN-}^\prime e^{-\lambda_{RN-}}- R_- \lambda_{RN-}^{\prime\prime}e^{-\lambda_{RN-}} + R_- \left (\lambda_{RN-}^\prime\right )^2 e^{-\lambda_{RN-}} \nonumber\\
&~ + \frac{1}{2} R_- e^{-\lambda_{RN-}} \left ( \xi^{RN}_0 - \xi^{RN}_1 \right )  \frac{1}{4} R_-^2 \lambda_{RN-}^\prime e^{-\lambda_{RN-}} \left ( \xi^{RN}_0 - \xi^{RN}_1 \right ) \nonumber \\
&~- + \frac{1}{4} R_-^2 e^{-\lambda_{RN-}} \left ( \xi^{RN\prime}_0 - \xi^{RN\prime}_1 \right ) + \frac{2A}{R_-^3} ~~~.\nonumber\\
\end{align}
Multiplying with $\frac{e^{\lambda_{RN_-}}}{R_-}$ and rearranging terms yields
\beqa
\frac{e^{\lambda_{RN_-}}}{R_-} \xi_2^{\prime} &= -\lambda^{\prime\prime}_{RN-} + \frac{1}{2} \left ( \xi^{RN}_0 - \xi^{RN}_1 \right ) + \frac{R_-}{4} \left ( \xi^{RN\prime}_0 - \xi^{RN\prime}_1 \right ) + \left (\lambda^{\prime}_{RN-} \right )^2 & \nonumber\\
&- \frac{1}{4} R_- \lambda^{\prime}_{RN-} \left ( \xi^{RN}_0 - \xi^{RN}_1 \right ) - \frac{2\lambda_{RN-}^{\prime}}{R_-} + \frac{2A}{R_-^4} e^{\lambda_{RN-}}
\eeqa
Hence we get equation (\ref{diffeqRN}) and we can use it for the following calculations.\\
At first we multiply the equation with $e^{-\lambda_{RN-}}$ and use (\ref{conxi}) to replace the $\xi^{RN}_\mu$. So the terms on the left side can be rewritten to
\begin{align}
\frac{\xi_2^{RN\prime}}{R_-} &= \frac{\xi_2^{\prime}}{R_-}\\
e^{-\lambda_{RN-}} \xi_1^{RN} &= e^{-\lambda_-} \xi_1
~~~,
\end{align}
whereas those on the right side transform to
\begin{align}
\lambda^\prime_{RN-} e^{-\lambda_{RN-}}(\xi_0^{RN} - \xi_1^{RN}) &= \lambda^\prime_{RN-} e^{-\lambda_-} (\xi_0 - \xi_1)\\
e^{-\lambda_{RN-}} (\xi_0^{RN\prime} - \xi_1^{RN\prime}) &= e^{-\lambda_{RN-}} \left ( e^{\lambda_{RN-} - \lambda_-} (\xi_0 - \xi_1)\right ) ^\prime\notag\\
&= (\lambda_{RN-}^\prime - \lambda_-^\prime) e^{-\lambda_-} (\xi_0 - \xi_1) \notag \\
&~~~ + e^{-\lambda_-} (\xi_0^\prime - \xi_1^\prime)\\
e^{-\lambda_{RN-}} (\xi_0^{RN} - \xi_1^{RN})^2 &= e^{\lambda_{RN-} - 2\lambda_-} (\xi_0 - \xi_1 )^2
~~~.
\end{align}
So we get
\begin{align}
\frac{\xi^\prime_2}{R_-} - e^{-\lambda_-}\xi_1 &= \frac{R_-}{4} \left (\lambda_{RN-}^\prime + \lambda_-^\prime \right ) e^{-\lambda_-} (\xi_0 - \xi_1)  \nonumber \\
&- \frac{R_-}{4} e^{-\lambda_-} (\xi_0^\prime - \xi_1^\prime ) - \frac{R_-^2}{8} e^{\lambda_{RN-} - 2\lambda_-} (\xi_0 - \xi_1 )^2 ~~~.\nonumber\\
\end{align}
Replacing $\xi_1$ with (\ref{xi1-xi2}) and subtracting (\ref{difa}) times $e^{-\lambda_-}$ leads to
\beqa
0 = \frac{R_-}{4} \left ( \lambda_{RN-}^\prime - \lambda_-^\prime \right ) e^{-\lambda_-} \left (\xi_0 - \xi_1 \right ) - \frac{R_-^2}{8} \left ( 1- e^{\lambda_{RN-} - \lambda_-} \right ) e^{-\lambda_-} \left ( \xi_0 - \xi_1 \right )^2 \nonumber\\
~~~.
\eeqa
Therefore we can conclude either
\beqa
\xi_0 - \xi_1 = 0 ~~~,
\eeqa
which combined with the ideal fluid ansatz and $p = w\rho$ leads to a case equivalent to introducing a cosmological constant, since (\ref{xik-rho}) demands $w = -1$.
Or the remaining part has to vanish which yields
\beqa
\xi_0 - \xi_1 = \frac{2\left ( \lambda_{RN-}^\prime - \lambda_-^\prime \right )}{R_- \left ( e^{\lambda_{RN-} - \lambda_-} -1 \right )} ~~~. \label{xi0-xi1RNvar}
\eeqa 
In the next step we want to eliminate all Reissner-Nordstr\"om variables. Therefore we begin with replacing $\lambda_{RN-}^\prime$ and for convience $\lambda_-^\prime$ as well
\begin{align}
\lambda_{RN}^\prime &= -\left (e^{-\lambda_{RN-}} \right )^\prime e^{\lambda_{RN-}} = -\left (e^{-\lambda_-} \right )^\prime e^{\lambda_{RN-}} - \frac{2A}{R_-^3} e^{\lambda_{RN-}}\nonumber\\
\lambda_-^\prime &= -\left (e^{-\lambda_-} \right )^\prime e^{\lambda_-}
~~~,
\end{align}
so that we can rewrite (\ref{xi0-xi1RNvar}) multiplied by $\frac{e^{-\lambda_-}}{2}$ to
\begin{align}
\frac{e^{-\lambda_-}}{2} (\xi_0 - \xi_1) &= \frac{- \left ( e^{-\lambda_-} \right )^\prime \left ( e^{\lambda_{RN-}} - e^{\lambda_-} \right ) - \frac{2A}{R_-^3} e^{\lambda_{RN-}}}{R_- \left ( e^{\lambda_{RN-}} - e^{\lambda_-} \right )} \nonumber\\
\frac{e^{-\lambda_-}}{2} (\xi_0 - \xi_1) &= - \frac{\left (  e^{-\lambda_-} \right )^\prime}{R_-} - \frac{2A}{R_-^4} \frac{e^{\lambda_{RN-}}}{\left ( e^{\lambda_{RN-}} - e^{\lambda_-} \right )}
~~~.
\end{align}
Now we rewrite the denominator of the second term
\begin{align}
e^{\lambda_{RN-}} - e^{\lambda_-} &= \frac{1}{e^{-\lambda_-} - \frac{A}{R_-^2}} - e^{\lambda_-} \nonumber\\
e^{\lambda_{RN-}} - e^{\lambda_-} &= \frac{1}{e^{-\lambda_-} - \frac{A}{R_-^2}} - \frac{1- \frac{A}{R_-^2 e^{\lambda_-}}}{e^{-\lambda_-} - \frac{A}{R_-^2}} = \frac{A}{R_-^2} e^{\lambda_-} e^{\lambda_{RN-}} ~~~, \nonumber\\
\end{align}
so that the equation is transformed into
\beqa
\frac{e^{-\lambda_-}}{2} (\xi_0 - \xi_1) = - \frac{\left (  e^{-\lambda_-} \right )^\prime}{R_-} - \frac{2}{R_-^2} e^{-\lambda_-} ~~~.
\eeqa

At this point we use the ideal fluid ansatz, we replace the $\xi$ according to (\ref{xik-rho}) and $e^{-\lambda_-}$ utilizing (\ref{g11})
\begin{align}
 \frac{8\pi\kappa}{c^2} \left ( \rho + \frac{p}{c^2} \right ) &= -\frac{2M_-}{R_-^3} - \frac{8\pi\kappa}{c^2R_-^3} \int R_-^2 \rho dR_- + \frac{8\pi\kappa}{c^2} \rho \notag \\
& ~~ - \frac{2}{R_-^2} + \frac{4M_-}{R_-^3} + \frac{16\pi\kappa}{c^2R_-^3} \int R_-^2 \rho dR_- \nonumber\\
\Rightarrow \frac{8\pi\kappa}{c^2} \frac{p}{c^2} &= - \frac{2}{R_-^2} + \frac{2M_-}{R_-^3} + \frac{8\pi\kappa}{c^2R_-^3} \int R_-^2 \rho dR_- ~~~.
\end{align}
Multiplying the equation with $\frac{c^2R_-^3}{8\pi\kappa}$ and differentiating it leads to
\beqa
3R_-^2 \frac{p}{c^2} + R_-^3 \frac{p^\prime}{c^2} = - \frac{c^2}{4\pi\kappa} + R_-^2 \rho 
\eeqa
Assuming the ansatz $\frac{p}{c^2} = w\rho$ we get
\beqa
w\rho^\prime = \frac{1-3w}{R_-} \rho - \frac{c^2}{4\pi\kappa R_-^3}  
\eeqa 
Now we have to distinguish between three cases. First of all let $w$ be 0, then we can easily calculate $\rho$
\beqa
\rho = \frac{c^2}{4\pi\kappa} R_-^{-2}~~~,
\eeqa
but with (\ref{add}) we observe, that $M_{de}$ is proportional to $R_-$, which leads to a constant correction within the metric
(see (\ref{e-lambda})). Thus this case is unphysical.\\
For the other two cases we have to solve the differential equation. This is easily done by calculating the homogeneous solution $\rho_h$
\beqa
\rho_h = CR_-^{\frac{1-3w}{w}}
\eeqa 
followed by a variation of the constant
\begin{align}
\rho &= C(R_-) R_-{\frac{1-3w}{w}}\\
\Rightarrow C^\prime &= -\frac{c^2}{4\pi\kappa w} R_-^{-\frac{1}{w}}
\end{align}
Let $w$ be 1, then $\rho$ is given by
\beqa
\rho = \tilde{C}R_-^{-2} - \frac{c^2}{4\pi\kappa} \ln \left ( \frac{R_-}{R_{0-}} \right ) R_-^2
~~~.
\eeqa
Again inserting this in (\ref{add}) excludes the case.\\
Now let us assume $w$ is any possible value except for 0, 1 and -1 (for w = -1 (\ref{xi0-xi1RNvar}) has not to be fullfilled, since $\xi_0 = \xi_1$) . In this case the variation of the constant leads to
\beqa
\rho = \tilde{C} R_-^{\frac{1-3w}{w}} - \frac{c^2}{4\pi\kappa \left ( w-1 \right )} R_-^{-2} ~~~.
\eeqa
Hence again $M_{de}$ has a term proportional to $R_-$, which however can be small, if the absolute value of w is much bigger than 1.
In this case we get in (\ref{e-lambda}) a correction almost of the order $\frac{1}{R_-}$, which is to high to be physical.
So we showed, that the ideal fluid ansatz is inconsistent with the assumption $\Xi^{RN}_\mu = \Xi_\mu$; at least with a linear correlation between $\frac{p}{c^2}$ and $\rho$.\\
Furthermore even if we would allow such corrections to the metric, the inconsistency of the ideal fluid ansatz with the assumption $\Xi^{RN}_\mu = \Xi_\mu$ can be shown. 
However to do that cumbersome calculations are needed, in which the solution for $\rho$ is inserted into the equation for $p^\prime$  
in (\ref{sum-eq}) and the orders of $R_-$ are compared.


\pagestyle{empty}
\renewcommand\theequation{A.\arabic{equation}}  
\setcounter{equation}{0}                                                                                                        
\end{document}